\documentclass[12pt]{iopart}

\newcommand{\lvec}{\mathbf{l}}
\newcommand{\nvec}{\mathbf{n}}
\newcommand{\mvec}{\mathbf{m}}

\newcommand{\rvec}{\mathbf{r}}

\usepackage{amssymb}

\usepackage{graphicx}

\begin{document}
                                   
\title{Superlight small bipolarons}

\author{J.P. Hague, P.E. Kornilovitch$^{\dagger}$, J.H. Samson and A.S. Alexandrov}
\address{Department
 of Physics, Loughborough University, Loughborough, LE11 3TU, United Kingdom}

\address{$\dagger$ Hewlett-Packard Company, 1000 NE Circle Blvd,
Corvallis, Oregon 97330, USA}

\date{\today}

\begin{abstract}

Recent angle-resolved photoemission spectroscopy (ARPES) has
identified that a finite-range Fr\"ohlich electron-phonon interaction
(EPI) with c-axis polarized optical phonons is important in cuprate
superconductors, in agreement with an earlier proposal by Alexandrov
and Kornilovitch. The estimated unscreened EPI is so strong that it
could easily transform doped holes into mobile lattice bipolarons in
narrow-band Mott insulators such as cuprates. Applying a
continuous-time quantum Monte-Carlo algorithm (CTQMC) we compute the
total energy, effective mass, pair radius, number of phonons and
isotope exponent of lattice bipolarons in the region of parameters
where any approximation might fail taking into account the Coulomb
repulsion and the finite-range EPI. The effects of modifying the
interaction range and different lattice geometries are discussed with
regards to analytical strong-coupling/non-adiabatic results. We
demonstrate that bipolarons can be simultaneously small and light,
provided suitable conditions on the electron-phonon and
electron-electron interaction are satisfied. Such light small
bipolarons are a necessary precursor to high-temperature Bose-Einstein
condensation in solids. The light bipolaron mass is shown to be
universal in systems made of triangular plaquettes, due to a novel crab-like motion. Another surprising
result is that the triplet-singlet exchange energy is of the first
order in the hopping integral and triplet bipolarons are heavier than
singlets in certain lattice structures at variance with intuitive
expectations. Finally, we identify a range of lattices where
superlight small bipolarons may be formed, and give estimates for
their masses in the anti-adiabatic approximation.


\end{abstract}

\maketitle

\section{Introduction}

A growing number of observations point to the possibility that
high-$T_{c}$ cuprate superconductors may not be conventional
Bardeen-Cooper-Schrieffer (BCS) superconductors \cite{bcs}, but rather
derive from the Bose-Einstein condensation (BEC) of real-space pairs,
as proposed by Mott and others
\cite{alemott94,aleedwards,alebook,edwards}. A possible fundamental
origin of such strong departure of the cuprates from conventional BCS
behaviour is the unscreened (Fr\"ohlich) EPI with a polaron shift, $E_p$ of
the order of 1 eV (La$_2$CuO$_4$, $E_p\approx0.65$eV) \cite{ale96},
routinely neglected in the Hubbard $U$ and $t-J$ models
\cite{tJ}. This interaction with $c-$axis polarized optical phonons is
virtually unscreened because the upper limit for the out-of-plane
plasmon frequency ($\lesssim 200$ cm$^{-1}$ \cite{plasma}) is well
below the characteristic frequency of optical phonons, $\omega\approx$
400 - 1000 cm $^{-1}$.  Since screening is poor, the magnetic
interaction remains small compared with the Fr\"ohlich EPI at any
doping of cuprates. In order to generate a convincing theory of
high-temperature superconductivity, one must treat the Coulomb
repulsion and \emph{unscreened} EPI on an equal footing. When both
interactions are strong compared with the kinetic energy of carriers,
the so-called ``Coulomb-Fr\"ohlich'' model (CFM) predicts a ground
state in the form of mobile, preformed, inter-site pairs dressed by
lattice deformations (i.e intersite bipolarons)
\cite{ale96,alekor,alebook}.

The most compelling evidence for (bi)polaronic carriers in novel
superconductors is the discovery of a substantial isotope effect on
the carrier mass \cite{guo} predicted by the bipolaron theory
\cite{aleiso}. Recent high resolution ARPES \cite{lan0,shen} provides
another piece of evidence for a strong EPI in cuprates between
electrons and c-axis-polarised optical phonons \cite{shen}. These, as
well as recent tunnelling \cite{tun}, earlier optical \cite{opt} and
neutron scattering \cite{neutron} experiments unambiguously show that
lattice vibrations play a significant though unconventional role in
high temperature superconductors.

Remarkably, earlier path-integral studies of large bipolarons in the
continuous limit \cite{ver}) led to a double surprise: (a) The large
bipolaron is only stable in a very limited sector of the parameter
space (Coulomb repulsion versus the Fr\"ohlich coupling constant) (b)
Most traditional ``Fr\"{o}hlich polaron'' materials (alkali halides
and the like) lie completely outside (and ``far'' from) this bipolaron
stability sector, but several high-$T_c$ superconductors lie very
close and even inside this rather restricted area of stability in the
parameter space.

When the strong Fr\"ohlich EPI operates together with a shorter range
deformation potential and molecular-type (e.g.  Jahn-Teller) EPIs, it
readily overcomes the Coulomb repulsion at short distances of about
the lattice constant, so that large (continuous) bipolarons become
local (lattice) bipolarons in narrow bands \cite{alebook}. Even at
significant doping local pairs are not overlapped, so that a high
critical temperature for Bose-Einstein condensation (BEC) could be
achieved, if they are sufficiently mobile. Analysis of the site-local
Holstein-Hubbard model has indicated that in order for the Coulomb
repulsion (Hubbard $U$) to be overcome by the induced attractive force
between electrons, EPI must be so large that the polaron (and
bipolaron) masses must be huge, rendering the transition temperature
minuscule \cite{aleran}. All is not lost, however, since the Holstein
interaction is the extreme short-range limit of a finite range EPI. Using the
finite-range EPI it is possible for electrons to pair between sites
\cite{ale96,alekor} without requiring the electron-phonon induced
attraction to be larger than the Hubbard $U$. Moreover, the individual
polarons are significantly lighter, so the mass of the pair has
potential to be orders of magnitude smaller than in the Holstein case.

To put these arguments on a solid microscopic ground we simulate the
CFM Hamiltonian on a lattice using an advanced QMC technique for
bipolarons and compare with analytic results in the strong coupling
and anti-adiabatic limits. First, we introduce the model.

\section{Coulomb-Fr\"ohlich model}

The Hamiltonian for the CFM is written as
%
\begin{eqnarray}
H = & - t \sum_{\langle \mathbf{nn'} \rangle\sigma}
c^{\dagger}_{\mathbf{n'}\sigma} c_{\mathbf{n}\sigma} + \frac{1}{2}\sum_{
\mathbf{nn'}\sigma\sigma'}V(\mathbf{n},\mathbf{n}') c^{\dagger}_{\mathbf{n}\sigma}
c_{\mathbf{n}\sigma}c^{\dagger}_{\mathbf{n'}\sigma'} c_{\mathbf{n'}\sigma'} \nonumber \\
 & +
\sum_{\mathbf{m}} \frac{\hat{P}^{2}_\mathbf{m}}{2M} +
\sum_{\mathbf{m}} \frac{\xi^{2}_{\mathbf{m}} M\omega^2}{2} -
\sum_{\mathbf{n}\mathbf{m}\sigma} f_{\mathbf{m}}(\mathbf{n})
c^{\dagger}_{\mathbf{n}\sigma} c_{\mathbf{n}\sigma} \xi_{\mathbf{m}}
\: . \label{eq:hamiltonian}
\label{FH}
\end{eqnarray}
%
Each ion has a displacement $\xi_\mathbf{m}$. Sites labels are
$\mathbf{n}$ and $\mathbf{m}$ for electrons and ions respectively. $c$
annihilate electrons. The phonons are Einstein oscillators with
frequency $\omega$ and mass $M$.
$\langle\mathbf{n}\mathbf{n}'\rangle$ denote pairs of nearest
neighbours, and
$\hat{P}_{\mathbf{m}}=-i\hbar\partial/\partial\xi_{\mathbf{m}}$ ion
momentum operators. The instantaneous interaction
$V(\mathbf{n},\mathbf{n}')$ has on-site repulsion $U$ and nearest
neighbour interaction $V$ (if the electron-phonon coupling term is
set to zero, one obtains the simple $UV$ model).  The force function
is of the screened Fr\"ohlich type,
\begin{equation}
f_{\mvec}(\nvec)=\frac{\kappa}{[(\mvec-\nvec)^2+1]^{3/2}}\exp\left(-\frac{|\mvec-\nvec|}{R_{sc}}\right)
\end{equation}
($\kappa$ is a constant) \cite{spencer}. We will also use a slightly
different notation for the electron-phonon interaction term here,
$H_{el-ph}=-\omega\sum_{ij\sigma}g_{ij}c^{\dagger}_{i\sigma}c_{i\sigma}(d^{\dagger}_j
+ d_j)$ where $g_{ij}$ is a dimensionless interaction proportional to
the force, and $d^{ \dagger}_j$ create phonons at site $j$. We set
$\hbar = 1$.

Such a model has a remarkable property. Unlike the site local Holstein
model, there is attraction (and potentially pairing) even in the presence of very strong on-site
Coulomb repulsion. The model is justified in the presence of
alternating planes of itinerant electrons and ions, where there is
strong screening along the $c$-axis.

\begin{figure}
\begin{indented}\item[]
\includegraphics[width = 115mm, height = 45mm]{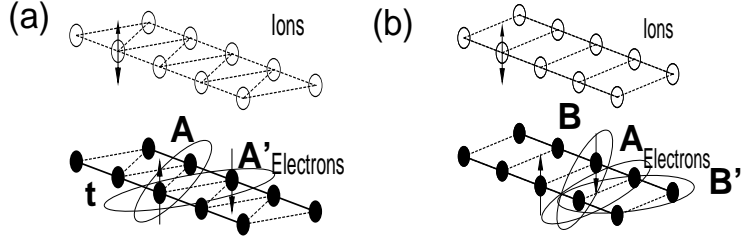}
\end{indented}
\caption{(a) Staggered vs (b) rectangular ladders. Ions are placed one
lattice spacing above a ladder of electrons, with one ion per
site. The ions are permitted to vibrate in the $z$-direction
only. Electrons inhabit one leg each, with no hopping between the
legs. In the strong coupling limit, there are significant geometrical
differences. On the staggered ladder, two degenerate near neighbour
pairs (A and A$'$) can form, which allows the polaron to scuttle in a
crab-like manner with mass proportional to the polaron
hopping. Alternatively, on the rectangular ladder there is one
near-neighbour state, and in order to move, the pair must break to
state A in order to change configuration from state B to B$'$. Such a state propagates by waddling awkwardly, and
has mass proportional to polaron hopping squared.}
\label{fig:laddersystems}
\end{figure}

There have been a number of studies discussing the masses of polarons
and bipolarons with long range interaction
\cite{korpolaron,hague,haguebipolaron}. The polaron formed from the
long-range Fr\"ohlich interaction proposed in \cite{ale96} has been
simulated in reference \cite{alekor}, demonstrating that the polaron
mass may be significantly lighter than its Holstein counterpart. This
is due to the nature of polarons in the Holstein case, which may be
demonstrated nicely by examining the Lang-Firsov transformation. In
that transformation, the operators in the Hamiltonian are replaced in
the following way,
\begin{eqnarray}
d^{\dagger}_j \rightarrow \tilde{d}^{\dagger}_j = d^{\dagger}_{j} + \sum_i g_{ij} n_i \label{eqn:lfone} & \hspace{10mm} &
c^{\dagger}_i \rightarrow \tilde{c}^{\dagger}_i = c^{\dagger}_i \exp[\sum_j g_{ij} (d^{\dagger}_j - d_j)] \\
d_j \rightarrow \tilde{d}_j = d_{j} + \sum_i g_{ij} n_i & \hspace{10mm} &
c_i \rightarrow \tilde{c}_i = c_i \exp[-\sum_j g_{ij} (d^{\dagger}_j - d_j)] \label{eqn:lffour}
\end{eqnarray}
thus hopping processes in the Holstein polaron (where $g_{ij} =
g\delta_{ij}$) take place by a complete relaxation of the lattice on
the initial site, a hop, and then a distortion on the target
site. With a longer range interaction, the lattice is pre-distorted
before the particle moves, leading to a much smoother process with a
lower intermediate energy state. We have recently determined that
long-range interactions lead to a reduction of the importance of
geometry on the properties of the bipolaron, especially the mass,
leading to very similar results on triangular and square lattice
\cite{hague}. We will discuss the crossover between Fr\"ohlich and
Holstein polarons later in this article.

The Lang-Firsov transformation is an exact canonical transformation,
and leads to a transformed Hamiltonian with a new transformed
wavefunction $|\Psi\rangle_{LF} = e^{-S}|\Psi\rangle$. It is most
instructive to consider the transformation of the atomic Hamiltonian
and the transformation of the hopping terms separately, since
typically the Lang-Firsov transformation is the starting point for a
series of perturbative analyses.

\subsection{Transforming the atomic Hamiltonian}

When the hopping term is set to zero, the phonon portion of the CFM is written as follows:
\begin{equation}
H_{at} = -\omega \sum_{ij} g_{ij} n_i (d^{\dagger}_j + d_j) + \omega \sum_j \left(d^{\dagger}_j d_j + \frac{1}{2}\right)
\end{equation}
(N.B. The index $i$ is now taken to contain a spin and a site index). Applying the Lang-Firsov canonical transformation \footnote{Inspection of equations \ref{eqn:lfone} and \ref{eqn:lffour} shows that electron number operators are unchanged on transformation, so the Coulomb part of the Hamiltonian is also unchanged},
\begin{eqnarray}
\tilde{H}_{at} & = & -\omega \sum_{ij} g_{ij} n_i (d^{\dagger}_j + d_j + 2\sum_{i'}g_{i'j}n_{i'}) \nonumber\\
& & + \omega \sum_j \left[(d^{\dagger}_j + \sum_{i}g_{ij}n_{i}) (d_j + \sum_{i'}g_{i'j}n_{i'}) + \frac{1}{2}\right]\\
& = & - \sum_{ii'} n_i n_{i'} \sum_{j} \frac{f_{ij} f_{i'j}}{2 \omega^2 M}  + \omega \sum_j \left(d^{\dagger}_j d_j + \frac{1}{2}\right)
\label{eqn:atomic_g}
\label{eqn:shift}
\end{eqnarray}
where we reordered the summation, and noted that
$g_{ij} = f_{ij} / \omega\sqrt{2M\omega}$.
This shows the remarkable property of the Lang-Firsov transformation,
since the electron and phonon subsystems in the atomic limit are now
completely decoupled. 

\begin{figure}
\begin{indented}\item[]
\includegraphics[height = 65mm, angle = 270]{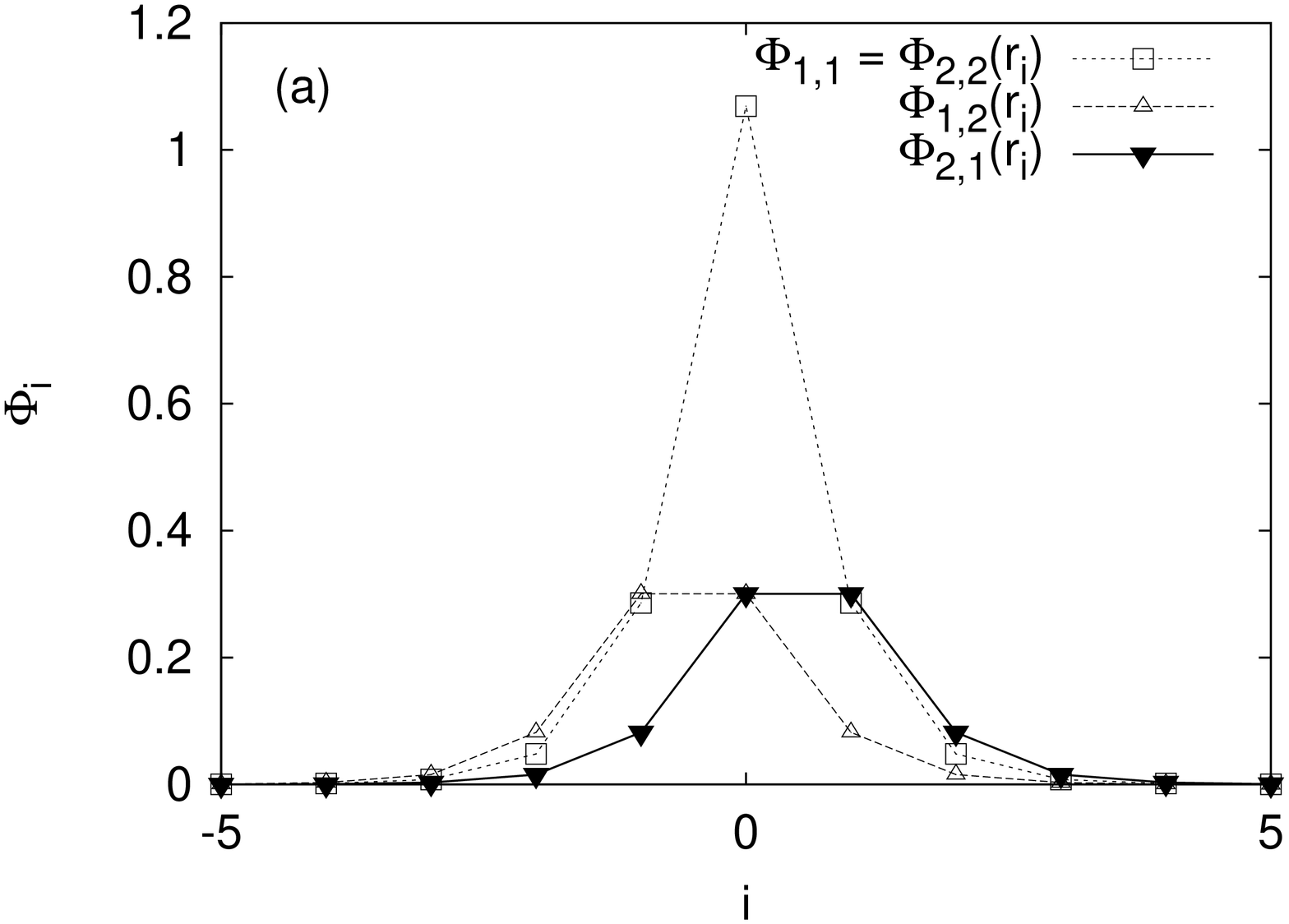}
\includegraphics[height = 65mm, angle = 270]{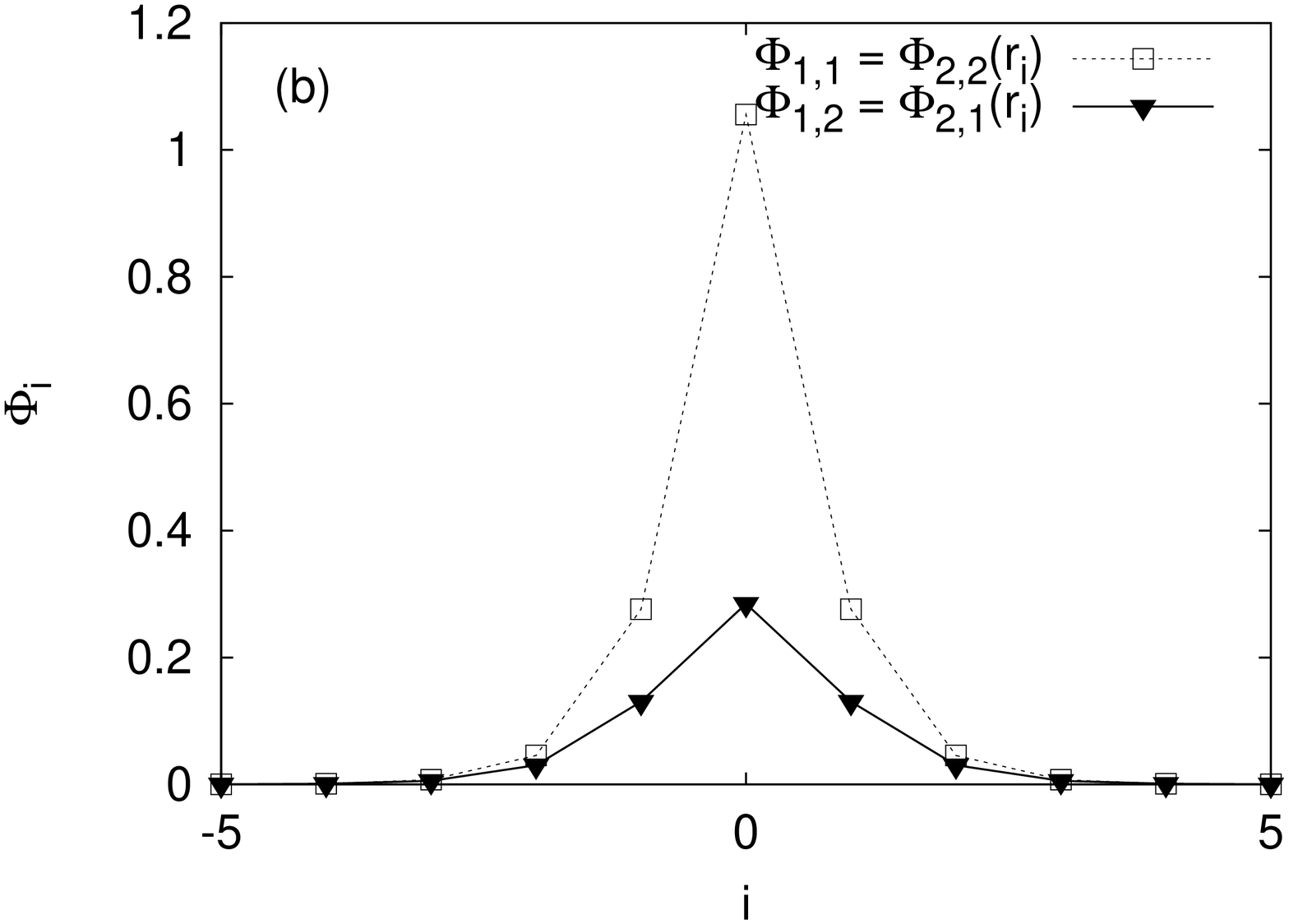}
\end{indented}
\caption{$\Phi$ functions for (a) the staggered ladder and (b) the
rectangular ladder (subscripts $1$ and $2$ correspond to the leg of the chain). Note that since an index is assigned to each
unit cell, there is an offset of 1 index in the interaction funcion
between paths 1 and 2, relative to the interaction function between
paths 2 and 1 when the staggered ladder is simulated.}
\label{fig:phifunction}
\end{figure}

At this stage, it is convenient to introduce the following function,
\begin{equation}
\Phi_{\Delta\mathbf{r}}[\mathbf{r}(\tau),
\mathbf{r}(\tau')]=\sum_{\mathbf{m}}f_{\mathbf{m}}[\mathbf{r}(\tau)]f_{\mathbf{m}+\Delta\mathbf{r}}[\mathbf{r}(\tau')]
\end{equation}
where the reason for adding an additional translation, $\Delta \rvec$,
to the phonon sub-system will become apparent when the action is
introduced in the next section. For the following discussion, $\Delta
\rvec=0$.  The $\Phi$ function for the ladder systems investigated in
this paper is shown in figure \ref{fig:phifunction}, corresponding to
a screened Fr\"ohlich interaction with $R_{sc} = 1$.
We also define the dimensionless interaction parameter $\lambda = E_p /
W$ where $W$ is the magnitude of the energy of the tight-binding electron (normally
the half band-width $zt$). As we can see from equation \ref{eqn:shift}, the energy shift when there is only one particle
(polaron shift for $i=i'$) is $E_p = \frac{1}{2M\omega^2}\sum_j f^2_{0j} =
\frac{\Phi_0(0,0)}{2M\omega^2}$. Thus $\lambda =
\frac{\Phi_0(0,0)}{2WM\omega^2}$. Substituting that definition into
the atomic Hamiltonian, one obtains:
\begin{eqnarray}
\tilde{H}_{at} & = & - \sum_{ii'} n_i n_{i'} \frac{W\lambda\Phi_0(i,i')}{ \Phi_0(0,0)}  + \omega \sum_j \left(d^{\dagger}_j d_j + \frac{1}{2}\right)
\end{eqnarray}
The reason for introducting the new functions $\Phi$ can immediately
be seen, since they appear in the Hamiltonian as a ratio. Thus, in
combination with $\lambda$ they give a universal definition of
coupling in models with long-range hopping.

\subsection{Transforming the electron hopping term}

Substitution of equations (\ref{eqn:lfone}-\ref{eqn:lffour})
transforms the tight binding Hamiltonian in the following way:
\begin{equation}
H_{tb} = \sum_{ii'} t_{ii'} c^{\dagger}_i c_{i'}
\rightarrow \tilde{H}_{tb} = \sum_{ii'} t_{ii'} \tilde{c}^{\dagger}_i \tilde{c}_{i'}
= \sum_{ii'} \sigma_{ii'} c^{\dagger}_i c_{i'}
\end{equation}
where
\begin{equation}
\sigma_{ii'} = t_{ii'}\exp\left(\sum_j g_{ij} (d^{\dagger}_{j} - d_{j})\right)\exp\left(-\sum_j g_{i'j} (d^{\dagger}_{j} - d_{j})\right)
\label{eqn:sigma}
\end{equation}
i.e. we can see that the electron-phonon interaction in the
transformed Hamiltonian is part of the hopping process, and that to
some extent, one may regard the operators $\tilde{c}$ as creating
polarons (i.e. electrons and a phonon cloud at the same time).

We now apply the following identity,
\begin{equation}
e^{A}e^{B}e^{-[A,B]/2} = e^{A+B} 
\label{eqn:exponentidentity}
\end{equation}
(which is valid if $C = [A,B]$ commutes with both $A$ and $B$) with
$e^{-A}e^{A} = 1$, which is always valid. Therefore,
%
$e^{A}e^{B} = e^{A+B} e^{[A,B]/2}$.
%
In equation \ref{eqn:sigma} we may choose $A = \sum_j g_{ij}
(d^{\dagger}_{j} - d_{j})$ and $B = -\sum_j g_{i'j}
(d^{\dagger}_{j} - d_{j})$. So,
%
$A+B = \sum_j (g_{ij} - g_{i'j}) (d^{\dagger}_j - d_j)$
and
$[A,B]  = 0$
%
Since the commutator is a number, equation \ref{eqn:exponentidentity} may be applied.

The hopping operator becomes:
\begin{equation}
\sigma_{ii'} = t_{ii'} \exp\left(- \sum_j (g_{ij} - g_{i'j}) d_j + \sum_j (g_{ij} - g_{i'j}) d^{\dagger}_j  \right)
\end{equation}

Now, the identity may be used again, making the grouping, $A = - \sum_j
(g_{ij} - g_{i'j}) d_j$ and $B = \sum_j (g_{ij} - g_{i'j})
d^{\dagger}_j$. Thus $[A,B] = -\sum_{j}
(g_{ij} - g_{i'j})(g_{ij} - g_{i'j})$, and
\begin{equation}
\sigma_{ii'} = t_{ii'} \exp\left[ -\frac{W\lambda}{\omega} \left(1 - \frac{\Phi_{0}(i,i')}{\Phi_{0}(0,0)} \right) \right] e^{ \sum_j (g_{ij} - g_{i'j}) d^{\dagger}_j } e^{ - \sum_j (g_{ij} - g_{i'j}) d_j }
\end{equation}

This form is particularly useful, due to the order of the creation and
annihilation operators. Thus, when one carries out the perturbation
theory, calculation is reduced to computing matrix elements of the
form, $\langle l | (\sum_{j} (g_{ij}- g_{i'j}) d^{\dagger}_j)^{n} (
\sum_{j'} (g_{ij'} - g_{i'j'})d_{j'})^{m} | l'\rangle$. It's
interesting to note that, when computed as an average over the atomic
wavefunction, the hopping integral may be regarded as being modified
by the EPI as $t_{ii'}\rightarrow t_{ii'}\exp\left( -W \lambda
\gamma_{ii'}/\omega \right) = \tilde{t}_{ii'}$, where $\gamma_{ii'} =
1-\Phi_0(i,i')/\Phi_0(0,0)$. Such an approximation is valid in the
anti-adiabatic limit, and will be revisited later in this paper. The band-narrowing factor was originally introduced by Tyablikov using an equations of motion scheme \cite{tyablikov}.

\section{Quantum Monte-Carlo simulation}

\begin{figure}
\begin{indented}\item[]
\includegraphics[width=90mm]{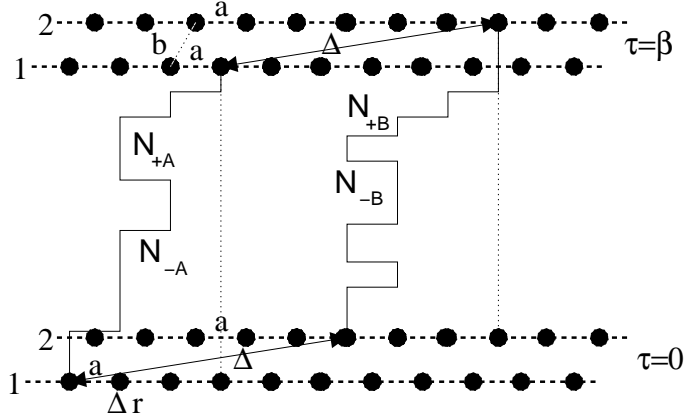}
\end{indented}
\caption{Example path configuration on the ladder system, showing the
notation used. Paths are separated by a vector $\mathbf{b}$, and sites
in the chain by $\mathbf{a}$. The path configuration at $\tau=0$ is
identical to that at $\tau=\beta$ up to a shift $\Delta r$ (i.e. the
ends of the beginning and end of the paths are separated by
$\Delta$). Each path has a number of kinks $N_{A/B +}$ and antikinks
$N_{A/B -}$}.
\label{fig:examplepath}
\end{figure}

The CTQMC algorithm presented here is an extension of a similar path-integral method for simulating the polaron problem \cite{korpolaron}. An integration over phonon degrees of freedom following Feynman leads to an effective action, which is a functional
of two polaron paths in imaginary time  which form the bipolaron and
is given by the following double integral when $\hbar\omega\beta\gg1$ \cite{haguebipolaron},
 %
\begin{eqnarray}
A[{\bf r}(\tau)] & = & \frac{z\lambda\bar{\omega}}{2\Phi_0(0,0)}
\int_0^{\bar\beta} \int_0^{\bar\beta} d \tau d \tau'
e^{-\bar{\omega} \bar\beta/2}  \sum_{ij}\Phi_0[\mathbf{r}_i(\tau),\mathbf{r}_j(\tau')] \nonumber\\
 & & \hspace{20mm}\times \left( e^{\bar{\omega}(\bar\beta/2-|\tau-\tau'|)} +
                               e^{-\bar{\omega}(\bar\beta/2-|\tau-\tau'|)} \right)\nonumber\\
 &+ & \frac{z\lambda\bar{\omega}}{\Phi_0(0,0)}
 \int_0^{\bar\beta} \int_0^{\bar\beta} d \tau d \tau' e^{- \bar{\omega} \tau}
 e^{-\bar{\omega}( \bar\beta - \tau')} \nonumber\\
&& \hspace{20mm}\times  \sum_{ij}\left( \Phi_{\Delta\mathbf{r}}[\mathbf{r}_i(\tau),\mathbf{r}_j(\tau')] -
 \Phi_0[\mathbf{r}_i(\tau),\mathbf{r}_j(\tau')]\right)\nonumber\\
&  - & \frac{1}{2}\int_0^{\beta}V(\mathbf{r}_1(\tau),\mathbf{r}_2(\tau))\,d\tau \: .
\label{eq:seven}
\end{eqnarray}
%
where the vector $\Delta\mathbf{r}=\mathbf{r}(\beta)-\mathbf{r}(0)$
is the difference between the end points of one of the paths in the
non-exchanged configuration (here $\bar{\omega}=\hbar \omega/t$  and
$\bar{\beta}=t/k_BT$).  The indices $i=1,2$ and $j=1,2$ represent
the fermion paths. $V(\mathbf{r}_1,\mathbf{r}_2)$ is an
instantaneous Coulomb repulsion. The part of the action depending on $\Delta {\bf r}$ arises because the entire phonon subsystem at $\tau=\beta$ must also be shifted when there is a shift in the electron sub-system between the start and end configurations. The definition of $\Delta{\bf r}$ and other nomenclature for the CTQMC simulation of ladder systems are shown in figure \ref{fig:examplepath}.

From this starting point, the bipolaron is simulated using the
Metropolis Monte-Carlo (MC) method. The electron paths are continuous
in time with hopping events (or kinks) introduced or removed from the
path with each MC step.  Analytic integration is performed over
sections of parallel paths.  The ends of the two paths at $\tau=0$ and
$\tau=\beta$ are related by an arbitrary translation,
$\Delta\mathbf{r}$. In contrast to the one-particle case, the fixing
of the end configurations limits the update procedure to inserting and
removing pairs of kinks and antikinks. We constrain particles to
opposite legs of the ladder, which corresponds to two species of
charged particles. In such a system, there is no exchange between
particles. Exchange and singlet triplet splitting from Quantum Monte-Carlo simulations is briefly discussed in this section, with an analytical discussion in section 5. A full discussion of the QMC procedure for exchange is left to a future article.

\subsection{Binary updates}

In the path integral QMC with two paths, it is necessary to make two kink
 operations simulataneously to ensure that the end configurations
 remain identical up to a translation. There exist two classes of
 update. First, kink/antikink pair addition/removal on a single path is
 useful, since it always maintains the same end configurations up to a
 change in the interpath displacement. Second, kink pair additions on
 different paths are needed in the analysis of the bipolaron mass,
 where the $\tau=\beta$ end configuration must be exactly equal to the
 $\tau=0$ end configuration up to a parallel shift. Within both
 subclasses, there are two specific update types that satisfy the
 imaginary-time boundary conditions, as follows:
\begin{itemize}
\item[(I)] Two kinks of the same type $\lvec$ are added to or removed from two different paths.
\item[(II)] A kink-antikink pair is added to or removed from one of the two paths. An antikink to kink $\lvec$ is a kink with
the opposite direction $-\lvec$.
\item[(III)] A kink of type $\lvec$ is inserted into one path, and another kink of the same type $\lvec$ is removed from the same path (kink shift).
\item[(IV)] A kink of type $\lvec$ is added to one path, and an
antikink $-\lvec$ is removed from the other path.
\end{itemize}

 An important property of the bipolaron system is that the type and
 time of added or removed kinks still does not define the new path
 unambiguously. Indeed, imagine a kink of type $\lvec$ being inserted
 on a single path at time $\tau_{\mathrm{ins}}$. This could change the
 path in two different ways: Either the path at times $\tau>\tau_{\mathrm{ins}}$
 is shifted in the direction $\lvec$, or the path at times
 $\tau<\tau_{\mathrm{ins}}$ is shifted in the antidirection
 $\lvec$. We refer to the former change as a top shift, and the latter
 as bottom shift. For the single-path (polaron) problem the
 distinction between the top and bottom shifts was not important
 because they are identical up to a translation of the entire
 path. This argument does not hold for two paths, since the resulting
 interpath distance in a binary update changes with shift type, thus a
 choice of shifts is an important part of the Monte Carlo update
 process. We proceed to derive the update probabilities for the
 Monte-Carlo scheme set out above.

There is considerable flexibility in choosing the probabilities for
adding and removing kinks. We choose an equal weighting scheme for choosing kinks, shifts and paths as follows:
\begin{itemize}
\item[Ai] Choose a kink type $\lvec$ from the list of all possible kinks,
with equal probability $P_{\lvec}=1/N_k$, where $N_k$ is the total
number of kink types. $P_{\lvec}$ cancels on both sides of all balance equations considered below.
\item[Aii] Anti-kink types are always determined from the kink type.
\item[Aiii] Shift type (top or bottom) is chosen with equal probability $P_s=1/2$
{\em independently} for the two kinks. $P_s$ also cancels on both sides of the balance equation.
\item[Aiv] Assign path A with equal probability $1/2$ from the two available paths.
\item[Av] Assign path B as the other path.
\end{itemize}
We will also choose kinks to add and remove according to the following
weightings, although there are some specific rules in the updates
below to deal with cases where there are no kinks or antikinks of the
chosen type on a path.
\begin{itemize}
\item[Bi] The probability density for kink time selection when adding a
kink is always $p(\tau) = 1/\beta$, ($0<\tau<\beta$).
\item[Bii] The probability for removing a kink of type $\lvec$ from path A in a configuration $C$ is $1/N_{A\lvec}(C)$ where $N_{A\lvec}(C)$ is the number of kinks of type $\lvec$ on the path A.
\end{itemize}

We note that the following is not the only
set of possible rules, however, we consider this to be the most transparent
method of choosing kinks to insert and remove.

\paragraph{ \label{sec:threeone}
Update type (I): Addition/removal of two kinks to or from different paths
}

Consider two configurations with two paths, $C$ and $D$, where configuration
$D$ has two more ${\bf l}$ kinks than $C$, one on the first path at
time $\tau_1$, and one on the second path at time $\tau_2$.  The
balance equation is
\begin{equation}
W(C) \cdot
Q_A(C)
\cdot P(C \rightarrow D) =
W(D) \cdot
Q_R(D)
\cdot P(D \rightarrow C)  \: .
\label{eq:sixteen}
\end{equation}
The relative weight of configurations $C$ and $D$ is $W(D)/W(C) =
(t_{\bf l} \Delta \tau)^2 e^{A(D)-A(C)}$.  In order to approach the
limit of continuous time, we rewrite the probability $Q_A$ of
selecting two kinks at $\tau_1$ and $\tau_2$ to add to different
paths given a configuration $C$ as $Q_{A}({\bf l}, \tau_1, A; \lvec,
\tau_2, B | C) = q_{A}({\bf l}, \tau_1, A; \lvec, \tau_2, B |
C)(\Delta \tau)^2$, where $q_{A}$ is the corresponding {\em
two-dimensional} probability density. The probability of selecting the
two specific kinks $\tau_1$ and $\tau_2$ to remove from the resultant
configuration ($D$) to get back to $C$ is a finite number $Q_{R}({\bf
l}, \tau_1, A; \lvec, \tau_2, B | D)$, which is the probability of
removing those two kinks given configuration $D$.  Substituting these
results into the balance equation, one may cancel the $(\Delta
\tau)^2$. Applying Metropolis, update probabilities are obtained as
follows:
\begin{equation}
P(C \rightarrow D) = \min \left\{ 1 \; ; \;
\frac{t^2_{\bf l}  Q_{\rm R}({\bf l}, \tau_1, A; \lvec, \tau_2, B | D)}
{q_{\rm R}({\bf l}, \tau_1, A; \lvec, \tau_2, B | C)}  e^{A(D)-A(C)} \right\}  \: ,
\label{eq:seventeen}
\end{equation}
\begin{equation}
P(D \rightarrow C) = \min \left\{ 1 \; ; \;
\frac{q_{\rm A}({\bf l}, \tau_1, A; \lvec, \tau_2, B | C)}
{t^2_{\bf l} Q_{\rm R}({\bf l}, \tau_1, A; \lvec, \tau_2, B | D) }
 e^{A(C)-A(D)} \right\}  \: .
\label{eq:eighteen}
\end{equation}
%
Thus we have obtained update probabilities that do not depend on the
time discretisation, and we can immediately take the limit of
continuous time. A similar approach can be taken for any of the update
types I to IV. We now demonstrate all steps in the derivation of the
first update probability as an example.

The rules and resulting probabilities are as follows:
\begin{enumerate}
\item Choose kink types, shifts and paths according to rules Ai-Av.
\item If the initial configuration has {\em at least} one ${\bf l}$ kink on
path A {\em and} on path B, then removal of a pair is proposed
with probability $P_{R}=1/2$, and addition of a pair is
proposed with probability $P_{A}=1/2$.  Otherwise, only pair
addition can be attempted and $P_A=1$.
\item If pair addition is selected, times are selected to insert one kink on path A, and another on path B with independent equal probability density (rule Bi).
\item If pair removal is chosen, then one candidate kink is selected with independent equal
probability from each of paths A and B in configuration $D$ (rule Bii).

\end{enumerate}
Implementing these choices, one obtains $q_{\rm A}({\bf l}, \tau_1, A; \lvec, \tau_2, B | C)= P_{\lvec} P_s^2 P_A(C)/\beta^2$ from the combination of rules (i), (ii), (iii). Likewise, the combination of rules (i), (ii) and (iv) specifies that $Q_{R}({\bf l}, \tau_1, A; \lvec, \tau_2, B | D) = P_{\lvec} P_s^2 P_R(D)/N_{\lvec A}(D)N_{\lvec B}(D)$.  Rule (iii) leads to $P_A(C) =1/2$ if $N_{A{\bf
l}}(C) \geq 1$ and $N_{B{\bf l}}(C) \geq 1$ and $P_A(C)=1$ otherwise. Since configuration $D$ always has sufficent kinks to make a removal, we always have $P_R(D) = 1/2$

leading to the
following acceptance rules:
%
\begin{equation}
P({\rm addition}) = P(C\rightarrow D)
= \min \left\{ 1 \; ; \;
\frac{P_R(D)(t_{\bf l} \beta)^2e^{A(D)-A(C)}} {P_A(C) N_{A{\bf l}}(D) N_{B{\bf l}}(D)}  \right\}
\end{equation}
\begin{equation}
P({\rm removal}) = P(D \rightarrow C)
= \min \left\{ 1 \; ; \;
\frac{P_A(C)N_{A{\bf l}}(D) N_{B{\bf l}}(D)}
{P_R(D)(t_{\bf l} \beta)^2e^{A(D)-A(C)}}  \right\}
\label{eq:twenty}
\end{equation}
Please note which configuration the kink numbers apply to. {\bf In the case
of kink addition, initial configuration is $C$ and final configuration
is $D$. In the case of kink removal, the initial configuration is $D$
and the final configuration is $C$.}

\paragraph{ \label{sec:threetwo}
Update type (II): Addition/removal of a kink-antikink pair to one path
}

The general properties of this update type are similar to update type
(I): the addition of a pair at times $\tau_1$ and $\tau_2$ is
characterized by a two-dimensional probability density $q_{\rm A}({\bf l}, \tau_1, A; -\lvec, \tau_2, A | C)$ while removal of a pair is
charaterized by a finite number $Q_{\rm R}({\bf l}, \tau_1, A; \lvec, \tau_2, A | D)$. Eqs.~(\ref{eq:seventeen})-(\ref{eq:eighteen}) still
apply with $t_{\bf l} t_{-{\bf l}}$ in place of $t^2_{\bf l}$.  Since
$t_{\bf l} = t_{-{\bf l}}$, the acceptance rules are idential, and
only $Q$ and $q$ differ.

We use the following rules.
\begin{enumerate}
\item Choose kink types, shifts and paths according to rules Ai-Av.
\item If the initial configuration has {\em at least} one ${\bf l}$ kink and one antikink ($N_{A\bf l} \geq 1$, $N_{A-{\bf l}} \geq 1$), then removal of a pair is proposed
with probability $P_R = 1/2$, and addition of a pair is proposed with probability
$P_A = 1/2$. Otherwise, addition
of a pair is proposed with probability $P_A = 1$.
\item If pair addition is selected, kink and antikink insertion times are selected with independent equal probability density for insertion on path A. (Rule Bi)
\item If pair removal is chosen, then one candidate kink and one candidate antikink are selected with independent equal
probability from path A in configuration $D$ (Rule Bii).
\end{enumerate}

One computes the update probabilities as before:
\begin{equation}
P({\rm addition}) = P(C\rightarrow D)
= \min \left\{ 1 \; ; \;
\frac{P_R(D)(t_{\bf l} \beta)^2e^{A(D)-A(C)}} {P_A(C) N_{A{\bf l}}(D) N_{A-{\bf l}}(D)}  \right\}
\label{eq:nineteen}
\end{equation}
\begin{equation}
P({\rm removal}) = P(D \rightarrow C)
= \min \left\{ 1 \; ; \;
\frac{P_A(C)N_{A{\bf l}}(D) N_{A-{\bf l}}(D)}
{P_R(D)(t_{\bf l} \beta)^2e^{A(D)-A(C)}}  \right\}
\end{equation}

\paragraph{ \label{sec:threethree}
Update type (III): Addition and removal of a kink to one path (kink shift)
}

This update type does not change the number of kinks, and hence does
not change the kinetic energy of the system.  We define a
configuration $C$, and a configuration $D$ which is identical to $C$
except that one kink has been shifted. To get from configuration $C$
to $D$, a kink is removed from path A at time $\tau_1$ and is
reinserted in the path at time $\tau_2$. Since $C$ and $D$ have equal
total kink number, the ratio of statistical weights is $W(D)/W(C) =
e^{A(D)-A(C)}$. From detailed balance and Metropolis:
%
\begin{equation}
P ( C \rightarrow D ) = \min \left\{ 1 \; ; \;
\frac{Q_{\rm R}({\bf l}, \tau_2, A|D)  q_{\rm A}({\bf l}, \tau_1, A|D)}
     {Q_{\rm R}({\bf l}, \tau_1, A|C)  q_{\rm A}({\bf l}, \tau_2, A|C)}
      e^{A(D)-A(C)} \right\}
\label{eq:thirty}
\end{equation}
There is only one update rule, since we can get from $D$ and $C$ using
exactly the same process as going from $C$ to $D$.  All attributes of
the kinetic energy have dropped out from the equations.  The
acceptance rules are determined solely by the electron-phonon
interaction, as expected for this update type.

In many practical situations it is reasonable to choose the functions $q$ and $Q$
independent of time $\tau$.  In this case the above expressions simplify
significantly.  In particular, consider the following set of update rules.
\begin{enumerate}
\item Choose kink types, shifts and paths according to the general rules Ai-Av.
\item If the path has no ${\bf l}$ kinks, $N_{\bf l} = 0$, the update attempt is aborted. Otherwise, $N_{\bf l} \geq 1$, so propose an $\lvec$ kink for removal with equal probability $1/N_{\bf l}(C)$ (rule Bii).
\item Propose time for new ${\bf l}$ kink with constant probability
density $p(\tau)=1/\beta$ (rule Bi).
\end{enumerate}

These rules result in cancelation of $Q_{\rm remove}$ and $q_{\rm add}$ from the
above equations, which reduce to
%
\begin{equation}
P (C \rightarrow D) =
\min \left\{ 1 \; ; \; e^{A(D)-A(C)} \right\}  \: ,
\label{eq:thirtytwo}
\end{equation}

\paragraph{ \label{sec:threefour}
Update type (IV): Kink addition to one path and antikink removal from the other path
}

In this update, a kink ${\bf l}$ is added to one path and an antikink $-{\bf l}$
is removed from the other path.  As a result, the $\beta$ end configuration shifts
as a whole by ${\bf l}$.  In a reciprocal process, a kink ${\bf l}$ is removed
from one path, an antikink $-{\bf l}$ is inserted in the other path, and the
$\beta$ configuration shifts by $-{\bf l}$.

The ratio of weights is $W(D)/W(C) = (t_{-{\bf l}}/t_{\bf l}) e^{A(D)-A(C)}$.
The balance equation is satisfied by the following solution
\begin{equation}
P ( C \rightarrow D ) = \min \left\{ 1 \; ; \;
\frac{  Q_{\rm R}({\bf l}; \tau_1; A | D)
                    q_{\rm A}  (-{\bf l}; \tau_2; B | D)}
       {   Q_{\rm R}(-\lvec; \tau_2; B | C)
                    q_{\rm A}   ({\bf l}; \tau_1, A | C)}
      e^{A(D)-A(C)} \right\}
\label{eq:thirtyfive}
\end{equation}
Since we can obtain the inverse process by changing kink $\lvec$ for its antikink, the update probability $P( D \rightarrow C)$ is not necessary since we always choose $\lvec$ from all kink types.

As the simplest implementation, the following rules are used,
\begin{enumerate}
\item Choose kink types, shifts and paths according to rules Ai-Av.
\item If path B has no antikinks $-{\bf l}$, $N_{B{-\bf l}} = 0$, then the
update attempt is aborted. Otherwise, $N_{B{-\bf l}} \geq 1$,
so an antikink is proposed for removal from path B with equal probability (rule Bii).
\item The time location for kink insertion on path A is proposed with
equal probability density (rule Bi).
 \end{enumerate}

With these assumptions, the acceptance probability takes the form:
\begin{equation}
P ( C \rightarrow D ) = \min \left\{ 1 \; ; \;
\frac{  N_{B -{\bf l}}(C)} {  N_{A{\bf l}}(D)}
      e^{A(D)-A(C)} \right\}  \: ,
\label{eq:thirtyseven}
\end{equation}

\subsection{Estimators}

When our Monte-Carlo scheme has reached equilibrium, we make a series of
measurements of physical properties. The ground state energy is:
\begin{equation}
\epsilon_0 = -\lim_{\beta\rightarrow\infty} \left[\left\langle \frac{\partial A}{\partial\beta} \right\rangle
- \frac{1}{\beta}\left\langle \sum_s N_s \right\rangle\right] \: ,
\label{eq:nine}
\end{equation}
where $N_s$ is the number of kinks of type $s$, and angular brackets denote ensemble
averaging.  The number of phonons is given by:
\begin{equation}
N_{\mathrm{ph}} = - \lim_{\beta\rightarrow\infty}\frac{1}{\bar{\beta}}
\left\langle \left. 
\frac{\partial A}{\partial \bar{\omega}}\right|_{\lambda\bar{\omega}}\right\rangle \: ,
\label{eq:ten}
\end{equation}
where the derivative is taken keeping $\lambda \bar\omega$ constant.  
The polaron band energy spectrum can be computed from:
\begin{equation}
\epsilon_{\bf k} - \epsilon_0= - \lim_{\beta\rightarrow\infty} \frac{1}{\beta} \ln
\langle \cos ({\bf k} \cdot \Delta{\bf r}) \rangle  \: ,
\label{eq:eleven}
\end{equation}
where ${\bf k}$ is the quasi momentum.  By expanding this expression in small
${\bf k}$, the $i$-th component of the inverse effective mass is obtained as
\begin{equation}
\frac{1}{m^*_i} = \lim_{\beta\rightarrow\infty}\frac{1}{\beta \hbar^2} \langle (\Delta {\bf r}_i)^2 \rangle \: .
\label{eq:twelve}
\end{equation}
Thus the inverse effective mass is the diffusion coefficient of the polaron
path in the limit of the infinitely long ``diffusion time'' $\beta$. The bipolaron radius is the average distance between paths,
\begin{equation}
R_{bp} =\left\langle\sqrt{\frac{1}{\beta}\int_{0}^{\beta}\Delta\mathbf{r}_{12}(\tau)^2
d\tau}\right\rangle
\end{equation}
Finally, the mass isotope coefficient, $\alpha_{m^*_i} = d \ln m^*_i / d \ln M$,
is calculated as follows
\begin{equation}
\alpha_{m^* _i}= \lim_{\beta\rightarrow\infty}\frac{\bar{\omega}}{2} 
\frac{1}{\langle(\Delta {\bf r}_i)^2 \rangle}
\left[\left\langle (\Delta {\bf r}_i)^2 
\left. \frac{\partial A}{\partial \bar{\omega}} \right\vert_{\lambda}
\right\rangle - \langle (\Delta {\bf r}_i)^2 \rangle
\left\langle \left. \frac{\partial A}{\partial \bar{\omega}} \right\vert_{\lambda}
\right\rangle \right] \: .
\label{eq:thirteen}
\end{equation}

\subsection{General Monte-Carlo considerations}

There are certain aspects of good practice for quantum Monte-Carlo simulations that we adhere to here. As always for Monte-Carlo
simulations a random number generator with sufficient period is
used. Measurements are performed every few steps to avoid unnecessary
correlations in results (the aim here is to spend no longer measuring
than simulating, since time correlated results do not make a large
contribution to more accurate measurement). Careful blocking analysis
with large blocking sizes $N_B$ is performed to determine accurate
error bars. To avoid anomalous error bars caused by long time
correlations, we compare error bars computed with two block sizes $N_B$ and $2N_B$.

\subsection{Exchanges}

Exchanges are significantly more complicated, with several possibilities for the exchange update involving inserting and removing kinks. In the exchanged configuration, there are an additional 4 update rules, and there is also an ambiguous configuration where both paths have the same start and end points, which leads to some small additional modifications. We defer a full discussion of exchange update rules to a later paper. On our ladder models, exchanges are not required, since electrons sit on opposite legs.

\subsection{Singlet-triplet splitting in the Monte-Carlo method}

A consequence of exchange is that singlet and triplet states are not
degenerate. We can see the singlet-triplet mass difference as a
consequence of interference between paths in the Monte Carlo
simulation.  We take a simplified one-dimensional example to
illustrate the mechanism.  Consider a bipolaron of separation $R$
propagating from the sites $\{0, R\}$ at $\tau=0$ to $\{\Delta r,
R+\Delta r\}$ at $\tau=\beta$.  We assume that the weight $w$ of a
single-electron path is a monotonically decreasing function of the
number $n$ of kinks in the path, and the paths with the smallest
number of kinks dominate.  This is likely to be valid in the
strong-coupling limit. We can therefore write the weight of as path as
a monotonically rapidly decreasing function $w(d)$, where $d$ is the
distance between endpoints.  We also neglect the interaction between
paths.

Consider first the case of periodic boundary conditions $\Delta r=0$.
The weights of the singlet and triplet bipolaron paths are
respectively the sum and the difference of the direct and exchange
paths.
\begin{eqnarray}
w_s(0) & \approx & w(0) w(0) + w(R) w(R), \label{sing0}\\
w_t (0) & \approx & w(0) w(0) - w(R) w(R) \label{trip0}.
\end{eqnarray}
Here if $w(R)\ll w(0)$, the singlet and triplet weights are dominated by the direct path and are nearly equal.  

Now consider a twist larger than the bipolaron radius, $\Delta r > R >0$.  The singlet and triplet weights are dominated by the shortest paths:
\begin{eqnarray}
w_s(0) & = & w(\Delta r) w(\Delta r) + w(\Delta r+R) w(\Delta r-R), \label{singr}\\
w_t (0) & = & w(\Delta r) w(\Delta r) - w(\Delta r+R) w(\Delta r-R) \label{tripr}.
\end{eqnarray}
In this case the total number of kinks in either the direct or the exchange path is the same ($2\Delta r$), so there will be substantial cancellation in the triplet case.  Thus the diffusion coefficient of the triplet bipolaron will be smaller, and the effective mass larger, than that of the singlet.

\section{Polarons on triangular and rectangular systems}

We briefly discuss the lattice dependent features of the
polaron problem here. For more detailed discussion of the polaron
problem, the reader is directed towards our papers on this subject
\cite{hague, spencer, korpolaron}, and to the paper by
Kornilovitch in this issue \cite{kornilovitch_mott}. In this
section, we specifically discuss the changes to the masses of polaron moving on trangular and square lattices as the screening length
$R_{sc}$ is varied. The polaron mass forms part of the argument later in this
article.

Screened Fr\"ohlich electron-phonon interactions were simulated by
Spencer {\it et al.} \cite{spencer}, who demonstrated a continuous
crossover between the Fr\"ohlich and Holstein limits on the chain. In
particular, the mass of the particle is found to be
light down to quite small screening radii, consistent with results by
Bon\v{c}a and Trugman \cite{aleran} for nearest-neighbour
electron-phonon interactions.

\begin{figure}
\begin{indented}\item[]
\includegraphics[height = 65mm, angle = 270]{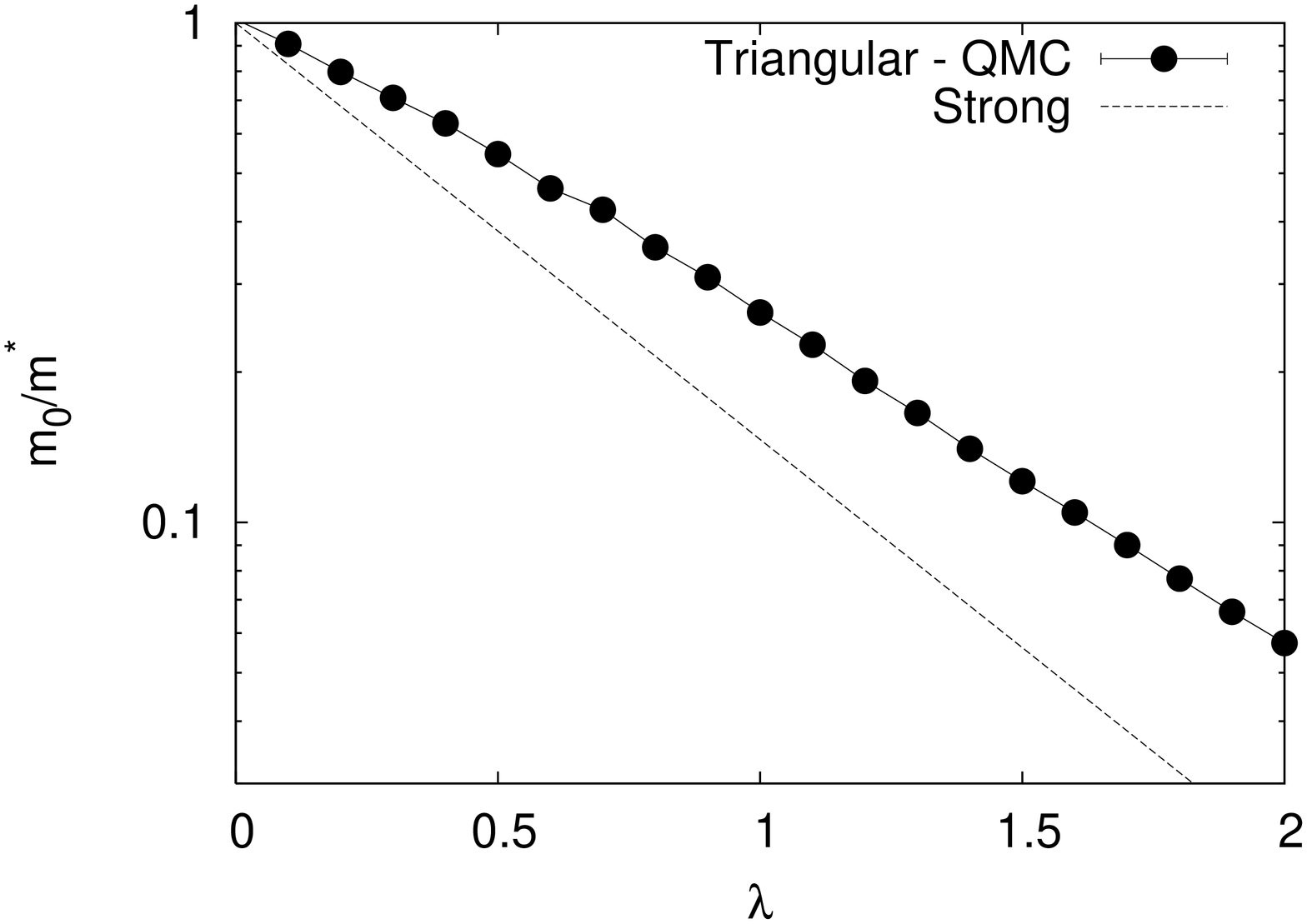}
\includegraphics[height = 65mm, angle = 270]{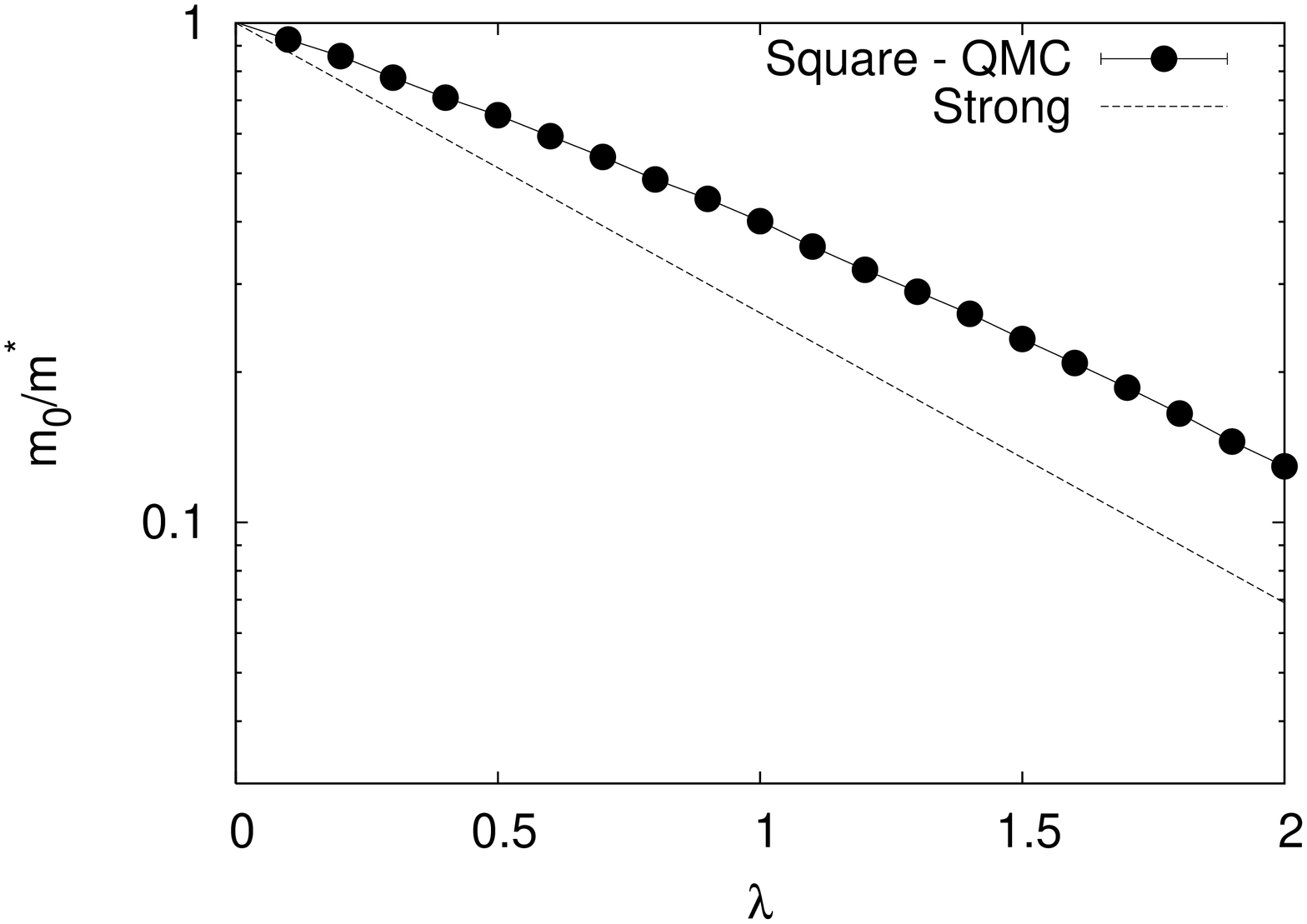}
\includegraphics[height = 65mm, angle = 270]{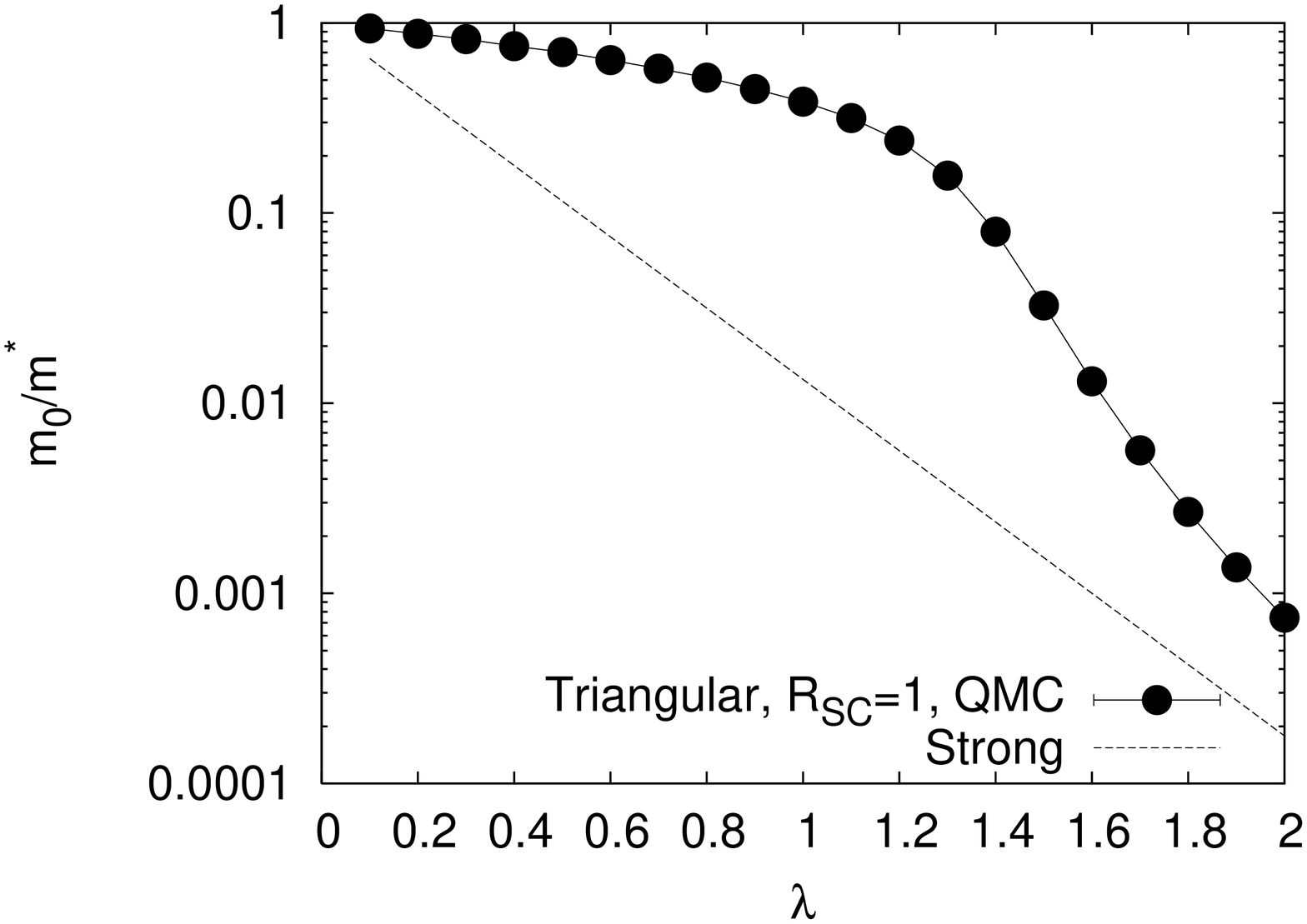}
\includegraphics[height = 65mm, angle = 270]{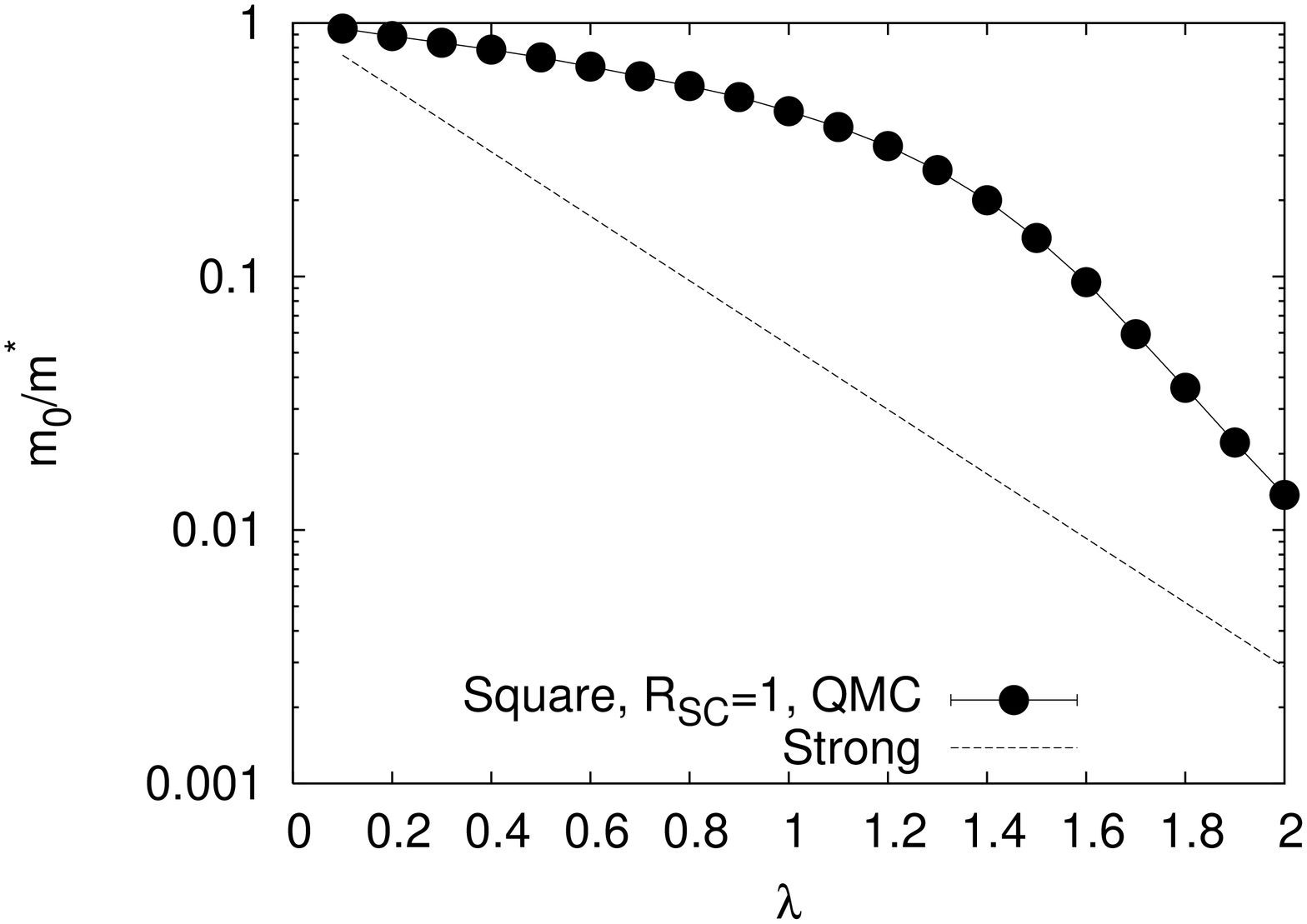}
\includegraphics[height = 65mm, angle = 270]{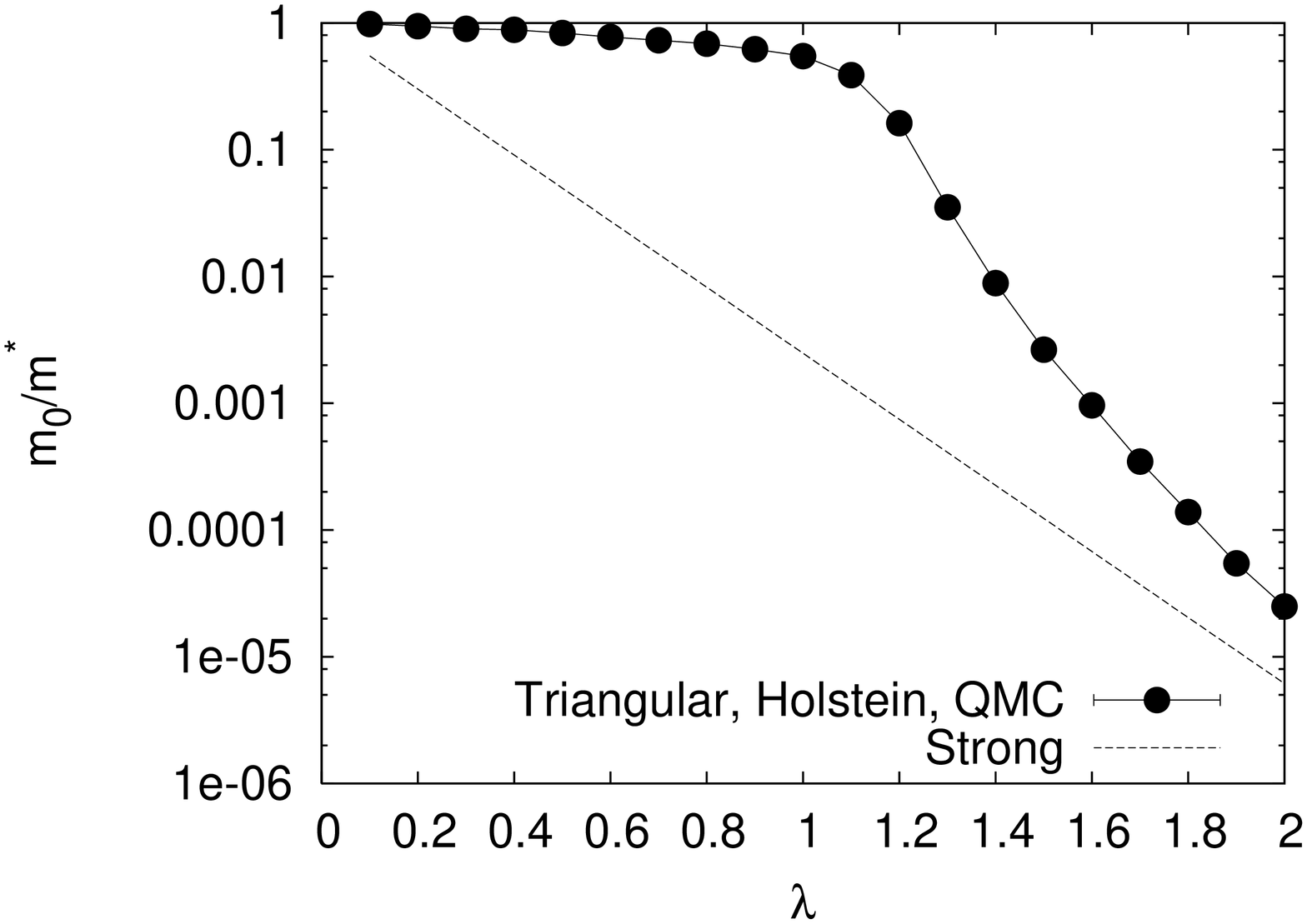}
\includegraphics[height = 65mm, angle = 270]{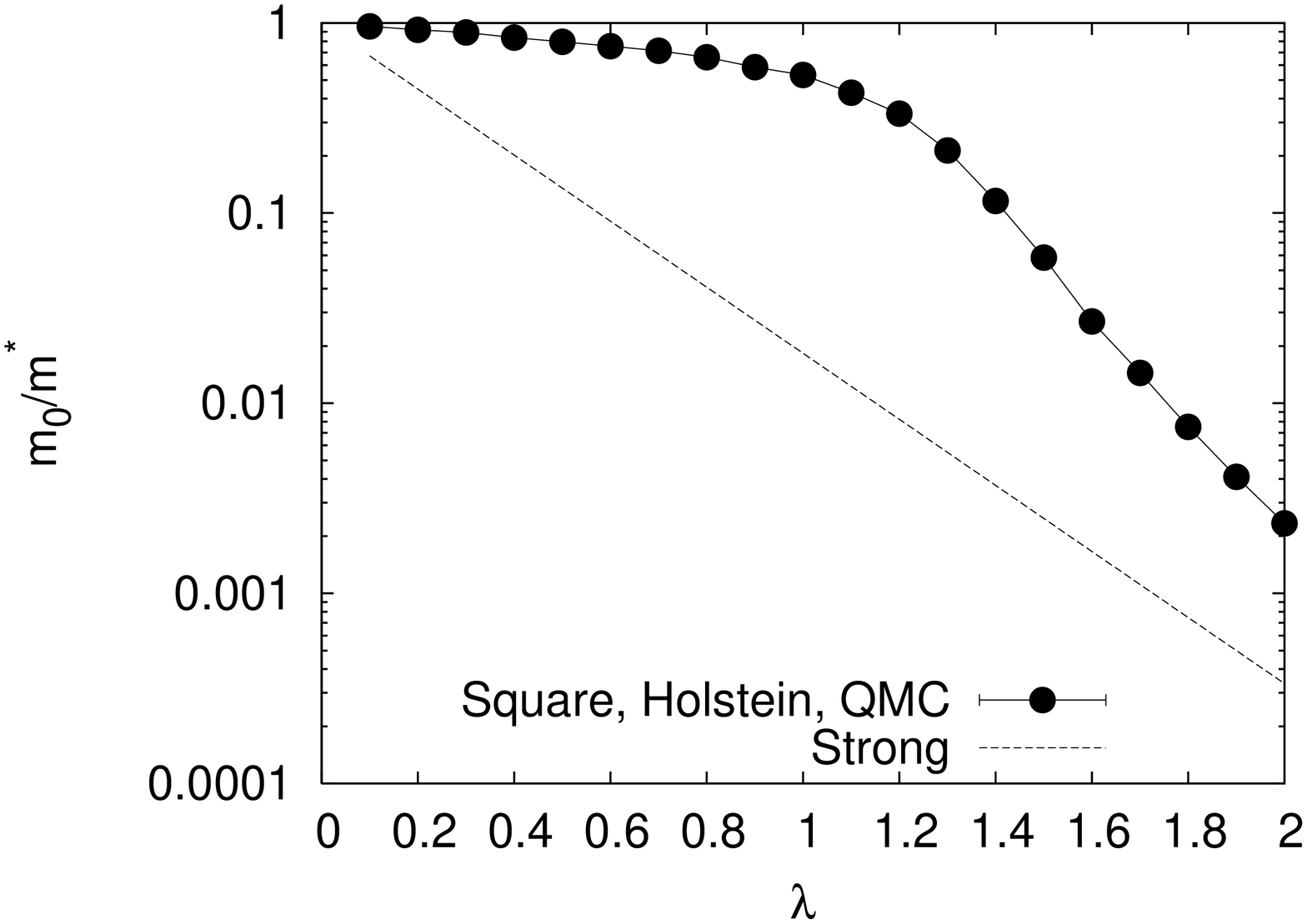}
\end{indented}
\caption{Inverse effective mass of the discrete Fr\"ohlich polaron on square
and triangular lattices.  $\omega/t = 1$, and $\lambda$ is
varied. Triangular lattices are shown on the left, and square lattices
are shown on the right. From top to bottom, the screening radius of
the interaction is decreased, with the top graphs showing Fr\"ohlich
interaction $R_{sc}\rightarrow\infty$, middle graphs screened interaction $R_{sc}=1$ and
the bottom graphs the Holstein interaction $R_{sc} = 0$. Fr\"ohlich polarons
are significantly lighter than their Holstein conterparts because the
long range interaction leads to pre-distortions of the lattice before a hopping.}
\label{fig:polaronmass}
\end{figure}

We have previously computed the properties of polarons on several
Bravais lattices, showing that the effects of the lattice type on the properties of the polaron are
`washed out' by long range interactions \cite{hague}. Figure
\ref{fig:polaronmass} shows the effective mass of the dicrete
Fr\"ohlich polaron on square and triangular lattices. Fr\"ohlich
polarons are significantly lighter than their Holstein counterparts,
due to the long range interaction. We have also shown that the
overriding factor for the properties of Holstein polarons is the
number of nearest neighbours in the lattice, and not the
dimensionality \cite{hague}. Since we are interested in long-range
interactions in this paper, then it suffices to note that the masses
of Fr\"ohlich polarons are of the same order of magnitude on all
lattice types, and that they are extremely light \cite{hague}. The
properties of realistic screened interactions lie somewhere inbetween,
but such polarons remain light down to quite small interaction ranges
of the order of a lattice spacing \cite{spencer}.

\section{Bipolarons on ladder systems}

The properties of bipolarons are significantly more complicated than
those of polarons. One expects bound Holstein bipolarons in
strongly correlated materials to be extremely heavy, since a very
large attractive potential is needed to overcome the repulsion. This
is not true for longer range interactions. According to work by
Bon\v{c}a and Trugman \cite{aleran}, even bipolarons with
nearest neighbour interactions are significantly more mobile than
their Holstein counterparts (indeed, one normally expects the mass of a
bipolaron to scale like the squared mass of the polaron, so polarons
which are an order of magnitude lighter than Holstein ones, may become
bipolarons which are two orders of magnitude lighter). Another
extremely interesting proposition is the role of geometry on the
bipolaron mass. When bipolarons are bound on nearest neighbour sites,
and another degenerate state may be reached in a single hopping event,
the leading correction to the atomic Hamiltonian is first order in the
hopping term, and not second order as one might expect for the
Holstein model, which leads to a bipolaron with a mass that is of
similar order of magnitude to the polaron mass (i.e. a superlight
bipolaron) \cite{alekor}. Such systems may be realised on triangular lattices, or on
lattices with large next-nearest neighbour hopping. We have recently
extended our Quantum Monte-Carlo algorithm to explore this type of
bipolaron, leading to similar conclusions \cite{haguebipolaron}. In
this paper, we discuss some of these extensions to the algorithm to
look at two types of ladder system shown in figure \ref{fig:laddersystems}.

\subsection{Weak and strong coupling}

Since the particles on the different legs of the ladder cannot
exchange the very weak-couping limit is not well bound, consisting of
two large polarons. As such, the weak-coupling perturbation theory can
be made about the unbound state. As will become apparent when we show
the quantum Monte-Carlo results, the number of phonons is small for
the weakly bound states, especially in the anti-adiabatic limit. The
perturbation theory in the electron-phonon coupling term only excites
single phonons, so if the number of phonons becomes too large, the
theory fails. In the presence of strong on-site repulsion, the bipolaron is not bound at zero coupling. In general, if there is a bound state for $\lambda\rightarrow 0^{+}$, then this perturbation expansion fails. The expansion is written as follows:
\begin{equation}
E_{\mathbf{k}} = 2\epsilon^{(0)}_{\mathbf{k}} -
2\frac{\lambda\omega W}{\Phi_0(0,0)}
\frac{1}{N} \sum_\mathbf{q}\frac{|f_{\mathbf{q}}|^2} {W(\mathbf{k},\mathbf{q})} \: ,
\label{eqn:wcenergy}
\end{equation}
\begin{equation}
W(\mathbf{k},\mathbf{q}) = 
\epsilon^{(0)}_{\mathbf{k}-\mathbf{q}} + \omega - \epsilon^{(0)}_{\mathbf{k}} \: ,
\end{equation}
\begin{equation}
f_{\bf q} = \sum_{\bf{m}} f_{\bf m}(0) e^{-i {\bf q} \cdot {\bf m}} \: ,
\end{equation}
where $N$ is the number of momentum states.  Thus the ground state polaron energy 
(at ${\bf k} = 0$) is $\epsilon^{(2)}_0 = -W + \lambda W \Gamma_{\epsilon_0}(\bar{\omega})$,
which defines a dimensionless coefficient $\Gamma_{E_0}$. 

There is no general analytic solution for the second order
perturbation theory, but values may be computed using numerical
integration. The number of phonons, isotope exponent and
inverse mass may also be written as, 
$N_{ph} = 2\lambda\Gamma_N(\hbar\omega)$, $\alpha =
2\lambda\Gamma_\alpha(\hbar\omega)$ and $2m_0/m^{**} =
1-\Gamma_m(\hbar\omega)\lambda$. The weak coupling limits for the
polaron are discussed in references \cite{spencer,hague} amoung
others. $\lambda$ is defined as before.

Aspects of the strong coupling limit are easy to compute from the path
integral formalism. In the very strong coupling limit, the most common
configurations are straight paths, since the action is proportional to
$\lambda$ and that configuration gives the smallest possible
action. Thus the strong coupling action reads:
\begin{equation}
A_{\rm strong} = \frac{\lambda W (\hbar\omega\beta - 1 + e^{-\hbar\omega\beta})}{\Phi_0(0,0)\hbar\omega}\sum_{ij}\Phi_0(i,j)
\end{equation}
For a polaron, the total energy for a self-interacting
strong-coupling system (i.e. a straight path) is $E=-W\lambda$, where
$W$ is the half band width, $-\epsilon_{\mathbf{k}=0}$. Computing the
action for the straight path configuration, one obtains the energy as
$E_{\rm strong}=-\partial A_{\rm strong}\partial\beta|_{\beta\rightarrow\infty}$. Leading to,
\begin{equation}
E_{\rm strong}^{\rm bp} = -2W\lambda[1+\Phi_0(0,\mathbf{b})/\Phi_0(0,0)]
\end{equation}
Where $\mathbf{b}$ is a nearest neighbour vector between chains
indicating the point of closest approach. This result is of little
suprise, since there are 2 polarons in a bipolaron, with
$E=-W\lambda$ each, and 2 inter-polaron binding terms with
$E=-W\lambda\Phi(0,\mathbf{b})/\Phi(0,0)$.
We may also compute the number of phonons associated with the
bipolaron in a similar manner using equation \ref{eq:ten} with
$A_{\rm strong}$, which we obtain as:
\begin{equation}
N^{\rm ph}_{\rm strong} = 2W\lambda[1+\Phi(0,\mathbf{b})/\Phi(0,0)]/\hbar\omega
\end{equation}

As we have discussed, the $\Phi$ functions for the ladder systems with
$R_{sc} = 1$ are plotted in fig. \ref{fig:phifunction}. Numerical
values for $\Phi(0,\mathbf{b})$, $\Phi(0,0)$ and
$1+\Phi(0,\mathbf{b})/\Phi(0,0)$ are shown in table \ref{tab:strong},
from which numerical values for the strong coupling behaviour are
computed. We discuss the region of validity of these results when we
compute the exact QMC results. In particular, if we plot appropriate
ratios, such as the ratio $N_ph\omega/\lambda$, we expect saturation
at strong coupling, consistent with table \ref{tab:strong}.

Typically, it has been discussed that an expansion of the Lang-Firsov
transformed Hamiltonian in the polaron hopping can be used at strong
coupling. The perturbation expansion to second order may be written as,
\begin{equation}
E_{tot} = E_{at} + \langle \Psi | \tilde{H}_{tb} |\Psi \rangle_{at} + \sum_{j} \frac{|\langle \Phi_j | \tilde{H}_{tb} | \Psi_{at} \rangle|^2}{E_j - E_{at}}
\end{equation}

The values for the strong coupling energy may be computed from the
leading term of this expansion, and have the same form as the energy
computed from the straight paths. This is not suprising, since the
corrections to the energy are at least as small as $\tilde{t}$.

There is a subtle point associated with phonon numbers. By
examination, the ground state of the phonon subsystem in the {\bf
transformed} Hamiltonian is $\sum_{j} d^{\dagger}_j d_j = \sum_{j} n_j
= 0$. In order to determine the total number of phonons in the true
ground state of the atomic Hamiltonian, one must transform back to the
regular wavefunction. One may either transform the wavefunction, or
the phonon part of the Hamiltonian. Transforming the phonon part of
the Hamiltonian is easy, and one obtains,
\begin{equation}
\langle \phi_{LF}| \tilde{H}_{ph} |\phi_{LF} \rangle = \langle \phi | H_{ph} |\phi \rangle = \omega N_{ph}
\end{equation}
thus
\begin{eqnarray}
N_{ph} & = & \langle \phi_{LF} | \sum_j (d^{\dagger}_j + \sum_{i}g_{ij}n_{i}) (d_j + \sum_{i'}g_{i'j}n_{i'}) |\phi_{LF} \rangle/\omega \\
& = & \langle \phi_{LF} | \sum_{ii'j} g_{ij} g_{i'j} n_i n_{i'} |\phi_{LF} \rangle/\omega
\end{eqnarray}
the expectation value of the phonons occupation may be rewritten in terms of $\lambda$ and $\Phi$, thus:
\begin{equation}
N_{ph} = -\frac{E_{at}}{\omega}
\end{equation}
(Note the minus sign in front of the equation - phonon number is positive).

Thus one determines that the total number of phonons in the bipolaron case is:
\begin{eqnarray}
N_{ph,at} & = &  \frac{2W\lambda}{\omega} \left(1 + \frac{\Phi_0(0,\mathbf{b})}{\Phi_0(0,0)}\right)
\label{eqn:atomicnph}
\end{eqnarray}
(N.B. A similar argument can be used for polarons.)

Unfortunately, the computation of the mass is not so simple. The mass
is very sensitive to the exact form of the renormalised hopping, and
the leading order of the perturbation expansion varies with lattice
type. For the superlight small bipolarons on triangular plaquettes discussed here, there is a
leading term with order $\tilde{t}$, and for rectangular systems, the
leading order is $\tilde{t}^2$. Examination of the path integrals
demonstrates that the mass is computed from a parallel shift. The
first perturbation of the paths from the straight line in the atomic
limit is the insertion of two parallel kinks (one on each path). In
order for the mass to be genuinely first order in the polaron hopping,
we would require updates with only one kink. Thus, there are
significant contributions from the second order term, including a
cancellation between the orders as $\lambda\rightarrow\infty$. In
fact, the expansion of the Lang-Firsov transformed Hamiltonian to 1st
order in the hopping turns out to correspond to the extreme
anti-adiabatic limit, which we now discuss.

\begin{table}
\begin{indented}\item[]
\begin{tabular}{llllll}
\hline
Lattice & $\Phi(0,0)$ & $\Phi(0,\mathbf{b})$ & $1+\frac{\Phi(0,\mathbf{b})}{\Phi(0,0)}$ & $E/\lambda$ & $\omega N_{ph}/\lambda$ \\
\hline
Staggered & 1.06896 & 0.30034 & 1.28096 & -5.12384 & 5.12384\\
Rectangular & 1.05554 & 0.284832 & 1.26984 & -5.07936 & 5.07936\\
\hline
\end{tabular}
\end{indented}
\caption{Strong coupling behaviour of the staggered and rectangular lattices.}
\label{tab:strong}
\end{table}

\subsection{Anti-adiabatic approximation}

The bipolaron dispersion can be evaluated analytically in the anti-adiabatic strong-coupling limit ($\hbar\omega\gg t, \lambda \gg 1$), in which case the problem can be reduced to that of a rigid dimer.
%
In this limit the Lang-Firsov transformation\cite{fir} eliminates the phonons from the Coulomb-Fr\"ohlich Hamiltonian (\ref{FH}). In combination with an averaging over phonons consistent with the anti-adiabatic approximation (if the phonon frequency is very large, there are no real phonons)  one obtains the following effective $UV$ Hamiltonian
%
\begin{eqnarray}
\label{Heff}
\tilde{H} &=& -\sum_{\mathbf{nn^\prime\sigma}}\tilde t_{\mathbf{nn^\prime}} c_{\mathbf{n\sigma}}^\dag  c_{\mathbf{n^\prime\sigma}}
-E_p\sum_{\mathbf{n\sigma}} c_{\mathbf{n\sigma}}^\dag  c_{\mathbf{n\sigma}} \nonumber \\ 
&+&\tilde U\sum_{\mathbf n} c_{\mathbf n\uparrow}^\dag c_{\mathbf n\uparrow} c_{\mathbf n\downarrow}^\dag  c_{\mathbf n\downarrow} 
+ \sum_{\mathbf{nn^\prime}} {}^\prime \sum_{\sigma\sigma^\prime}\tilde V_{\mathbf{nn^\prime}} c_{\mathbf{n}\sigma}^\dag  c_{\mathbf{n}\sigma} c_{\mathbf{n^\prime}\sigma^\prime}^\dag  c_{\mathbf{n^\prime}\sigma^\prime}.
\end{eqnarray}
where the primed sum excludes self-interaction, and the renormalised on-site and inter-site interactions are
\begin{eqnarray}
\label{Utilde}
\tilde U &=& \frac{V(\mathbf 0, \mathbf 0)}{2} - W\lambda
\end{eqnarray}
and
\begin{eqnarray}
\label{Vtilde}
\tilde V_{\mathbf{nn^\prime}}
&=&\frac{V(\mathbf{n, n^\prime})}{2}-\frac {W\lambda\Phi_\mathbf 0 (\mathbf n, \mathbf {n^\prime})} {\Phi_\mathbf 0 (\mathbf 0, \mathbf 0)}
\end{eqnarray}
respectively.  (There is no retardation in the anti-adiabatic limit.)
 Now suppose the on-site repulsion $\tilde U\gg \tilde t$ is large and repulsive and the inter-site interaction has a well-defined minimum $V_\mathrm{min}<0$ at some separation.  Let $N_\mathbf n$ be the set of sites of $\mathbf n^\prime$ at that separation from $\mathbf n$:
\begin{equation}
\label{Ni}
N_\mathbf n = \{{\mathbf n^\prime}: \tilde V_{\mathbf{nn^\prime}} =V_\mathrm{min} \}
\end{equation}
and
\begin{equation}
\label{notNi}
\tilde V_{\mathbf{nn^\prime}} - V_\mathrm{min} \gg \tilde t \;\mathrm{for}\; \mathbf n^\prime \notin N_{\mathbf n}.
\end{equation}
We shall call $N_\mathbf n$ the  ``neighbours'' of $\mathbf n$.  In general, $N_\mathbf n$ need not be the hopping neighbours  $\{{\mathbf n^\prime}: \tilde t_{\mathbf{nn^\prime}} \ne 0 \}$. 
The low-energy sector for two electrons corresponds to dimer states (bipolarons) in which the electrons are on neighbouring sites; the energies in this sector are $V_\mathrm{min} + O(\tilde t)$.  The energy cost of internal excitations of the bipolarons introduces a gap.

We can now sharply distinguish two types of bipolaron motion:
``crab-like'', in which the constituent polarons remain neighbours,
and ``crawler'', which requires virtual transitions out of the
low-energy sector.  The crab-like bipolaron bandwidth will be
$O(\tilde t)$, while the crawler contributions will be $O(\tilde
t^2)$. We shall for the present purposes project onto the low energy
sector and hence ignore the resulting higher-order terms in $\tilde
t$, immobilising crawlers. For simplicity we drop the tildes from the
notation and absorb the polaron shift $-E_p$ into the chemical
potential.

If a lattice $\Lambda$ has $L$ sites with a mean number $\nu$ of
neighbours per site, then the single-polaron Hilbert space is
$2L$-dimensional, the two-polaron Hilbert space is $4L^2$-dimensional
and the crab bipolaron Hilbert space is just $4\nu L$-dimensional.  We
can further reduce to one singlet and three triplet spaces, each of
dimensionality $\nu L$.  The singlet bipolaron space is

\begin{equation}
\label{singlet}
\mathcal S = \mathrm{span} \left\{
\frac 1 {\sqrt 2}\left(|\mathbf n\uparrow \mathbf{n^\prime}\downarrow\rangle + |\mathbf{n^\prime}\uparrow \mathbf n\downarrow\rangle\right): \mathbf n\in \Lambda, \mathbf{n^\prime}\in N_\mathbf n\right\}
\end{equation}
and the $S_z=1,0,-1$ sectors of the triplet bipolaron space are
\begin{eqnarray}
\mathcal T _1&=& \mathrm{span} \left\{
|\mathbf n\uparrow \mathbf{n^\prime}\uparrow\rangle: \mathbf n\in \Lambda, \mathbf{n^\prime}\in N_\mathbf n\right\}\label{triplet1}\\
\mathcal T _0&=& \mathrm{span} \left\{
\frac 1 {\sqrt 2}\left(|\mathbf n\uparrow \mathbf{n^\prime}\downarrow\rangle - |\mathbf{n^\prime}\uparrow \mathbf n\downarrow\rangle\right): \mathbf n\in \Lambda, \mathbf{n^\prime}\in N_\mathbf n\right\} \label{triplet0} \\
\mathcal T _{-1}&=& \mathrm{span} \left\{
|\mathbf n\downarrow \mathbf{n^\prime}\downarrow\rangle: \mathbf n\in \Lambda, \mathbf{n^\prime}\in N_\mathbf n\right\}.\label{triplet-1}
\end{eqnarray}

This enables us to write the low-energy effective Hamiltonian of the
dimers in each sector as a tight-binding Hamiltonian on the
\emph{dimer lattice} constructed in the following way: a node is
placed on the line joining neighbours in the lattice $\Lambda$.  If
$\mathbf{n^\prime}$ and $\mathbf{n^{\prime\prime}}$ are both hopping
neighbours of $\mathbf n$
($\tilde{t}_{\mathbf{n^\prime}\mathbf{n^{\prime\prime}}}\ne 0$), then
the dimer can hop from $\mathbf{n}\mathbf{n^{\prime}}$ to
$\mathbf{n}\mathbf{n^{\prime\prime}}$.  A bond is then drawn between
the two nodes on the dimer lattice with hopping integral
$\tilde{t}_{\mathbf{n^\prime}\mathbf{n^{\prime\prime}}}$ in the
singlet sector and
$-\tilde{t}_{\mathbf{n^\prime}\mathbf{n^{\prime\prime}}}$ in the
triplet sector. This sign change ensures the correct exchange symmetry
for closed paths on odd-membered rings: as a dimer completes one cycle
of an odd-membered ring its end-points are interchanged.  This can
lead to a dramatic difference between singlet and triplet bipolaron
masses on a non-bipartite lattice, as we shall see.

\subsection{Ladders}

In the staggered ladder depicted in Fig \ref{fig:laddersystems} the
neighbours of a site on one chain are the two adjacent sites on the
opposite chain, while the hopping neighbours of a site are along the
same chain. The corresponding dimer lattice is a one-dimensional chain
with hopping $\pm \tilde{t}$ and two sites per unit cell.  As there
are no exchange paths, the sign of the hopping can be gauged away and
there is no singlet-triplet splitting.  The polaron and bipolaron
dispersions are therefore respectively
\begin{eqnarray}
E_\mathrm {pol}(k) & = & -\tilde{t}\cos ka, \label{Epladder}\\
E_\mathrm {bip}(k) & = & \pm \tilde{t}\cos \frac{ka}{2}. \label{Ebladder}
\end{eqnarray}
The bipolaron effective mass is therefore four times that of the polaron (as previously reported \cite{alekor}):
\begin{eqnarray}
\label{mpladder}
m^* &=& \hbar^2 \left.\frac{d^2 E_{\rm pol}(k)}{dk^2}\right|_{k=0} = \frac{\hbar^2}{2\tilde{t}a^2}, \\
\label{mbladder}
m^{**} &=& \hbar^2 \left.\frac{d^2 E_{\rm bp}(k)}{dk^2}\right|_{k=0}  = \frac{2\hbar^2}{\tilde{t}a^2}.
\end{eqnarray}

This result is remarkable if one considers the standard strong coupling result for the mass of the bipolaron. In the rectangular ladder the requirement for a virtual internal excitation of the bipolaron in the crawler dynamics would lead to a hopping $O(\tilde{t}^2)$ and hence mass $O((m^{*})^2)$. The staggered ladder with long range electron phonon attraction has two degenerate nearest neighbour bound states (as summarised in figure 12), so no intermediate state is required when the particle hops. It is clearly important that the electrons are bound one lattice spacing apart to take advantage of this effect, but this can easily be achieved in the presence of a strong site-local Coulomb repulsion (so that the energy is not at a minimum when both particles are on the same site). This spaced minimum is clearly extremely important when real lattices beyond the toy ladder models presented here are considered. Some details of real lattices will be discussed later in this paper, but to briefly summarise, in order to obtain this special kind of bipolaron, (a) The lattice must have several degenerate near-neighbour bound states which can be transformed from one degenerate state to another via a single hop (b) Strong Coulomb repulsion is required to stop a unique single-site bound state from forming between the polarons, and (c) long range attraction is required so that the minimum in the potential function (attraction + repulsion) is at approximately one lattice spacing. This information is summarised in figures \ref{fig:pictorial} and \ref{fig:effinteraction} in the conclusion to this paper.

\subsection{QMC}

\begin{figure}
\begin{indented}\item[]
\includegraphics[height = 65mm, angle = 270]{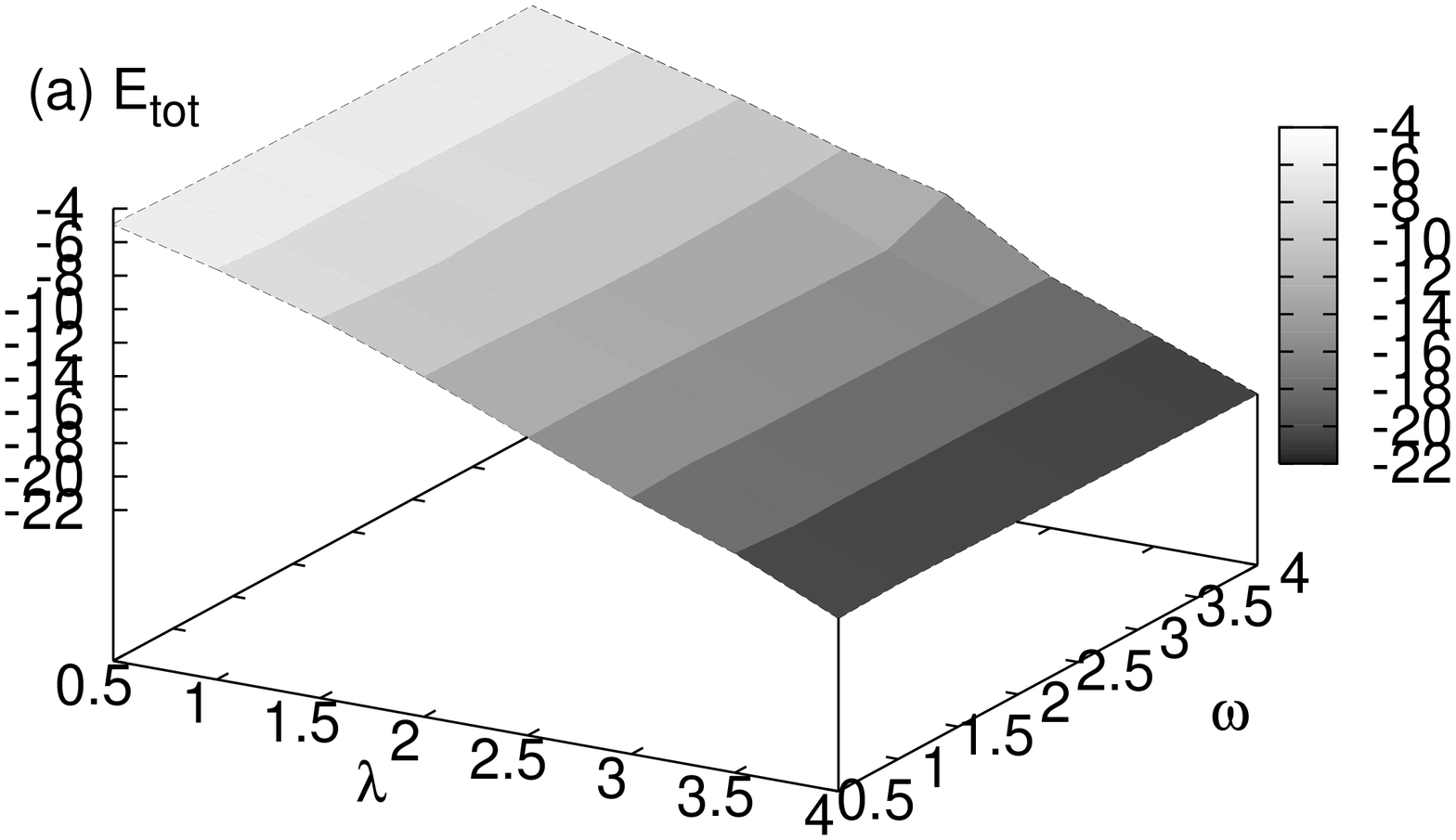}
\includegraphics[height = 65mm, angle = 270]{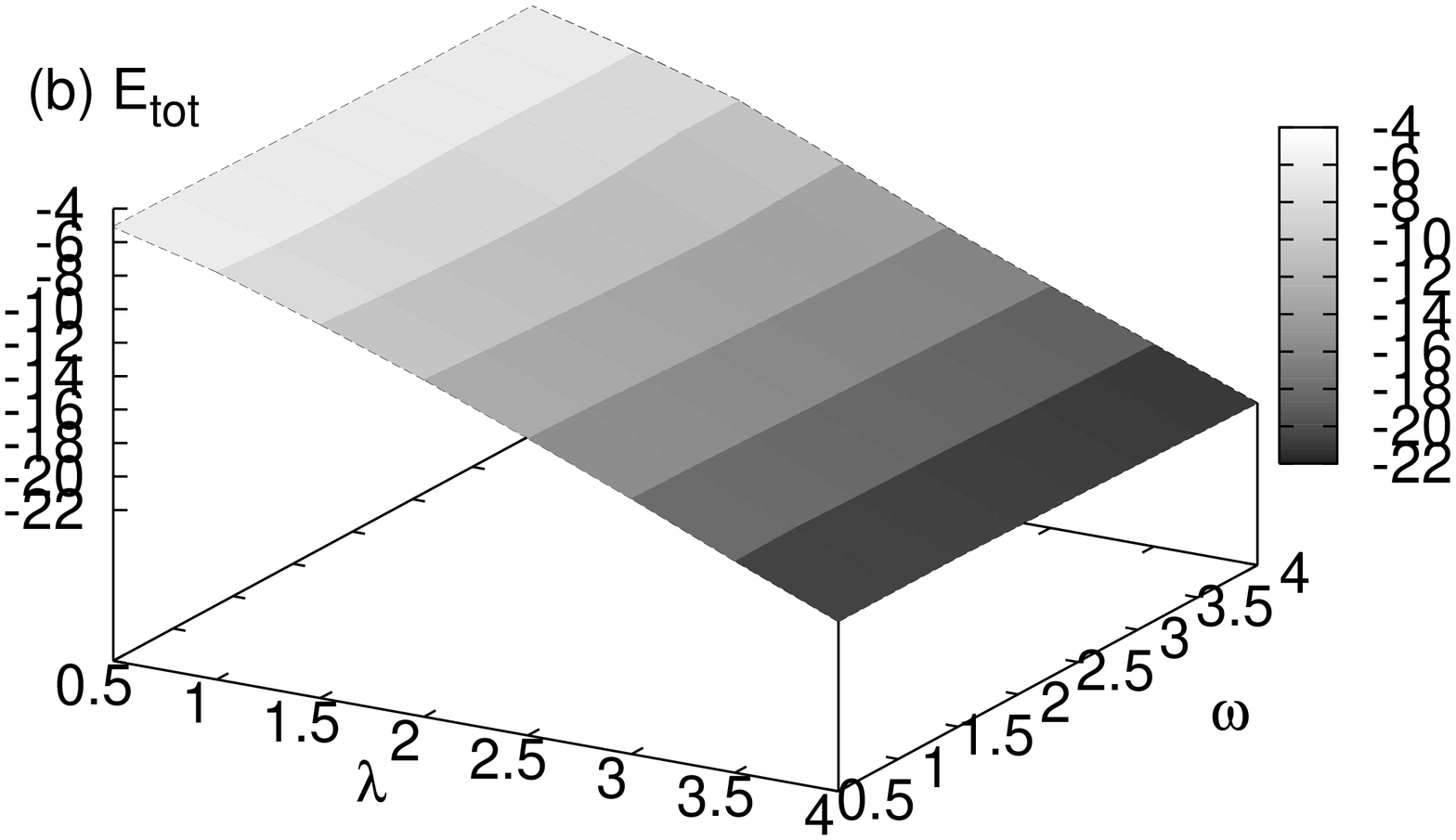}
\end{indented}
\caption{Total energy of the bipolaron for (a) the rectangular ladder
and (b) the staggered ladder. A slight change in gradient between weak
and strong coupling is just discernable, demonstrating (as we shall
see in the coming figures) that the staggered ladder reaches the
strong coupling limit at significantly lower $\lambda$ than the rectangular
ladder.}
\label{fig:totalenergy}
\end{figure}

We compute QMC results for bipolarons moving on staggered and
rectangular ladders with period 1000 for a range of $\lambda$ and $\omega$, including
the total energy, inverse bipolaron mass, bipolaron radius, mass
isotope exponent and the number of phonons in the bipolaron
cloud. Figure \ref{fig:totalenergy} shows the total energy of the
bipolaron for (a) the rectangular ladder and (b) the staggered
ladder. A slight change in gradient between weak and strong coupling
is just discernable, demonstrating (as we shall see in the coming
figures) that the staggered ladder reaches the strong coupling limit
at significantly lower $\lambda$ than the rectangular ladder. Strong
coupling results from the previous section agree well. We see that
there are no significant differences between the total energy of the
bipolarons on the staggered and rectangular ladders, although the
strong coupling limit is reached for slightly lower $\lambda$ in the
case of the staggered ladder.

\begin{figure}
\begin{indented}\item[]
\includegraphics[height = 65mm, angle = 270]{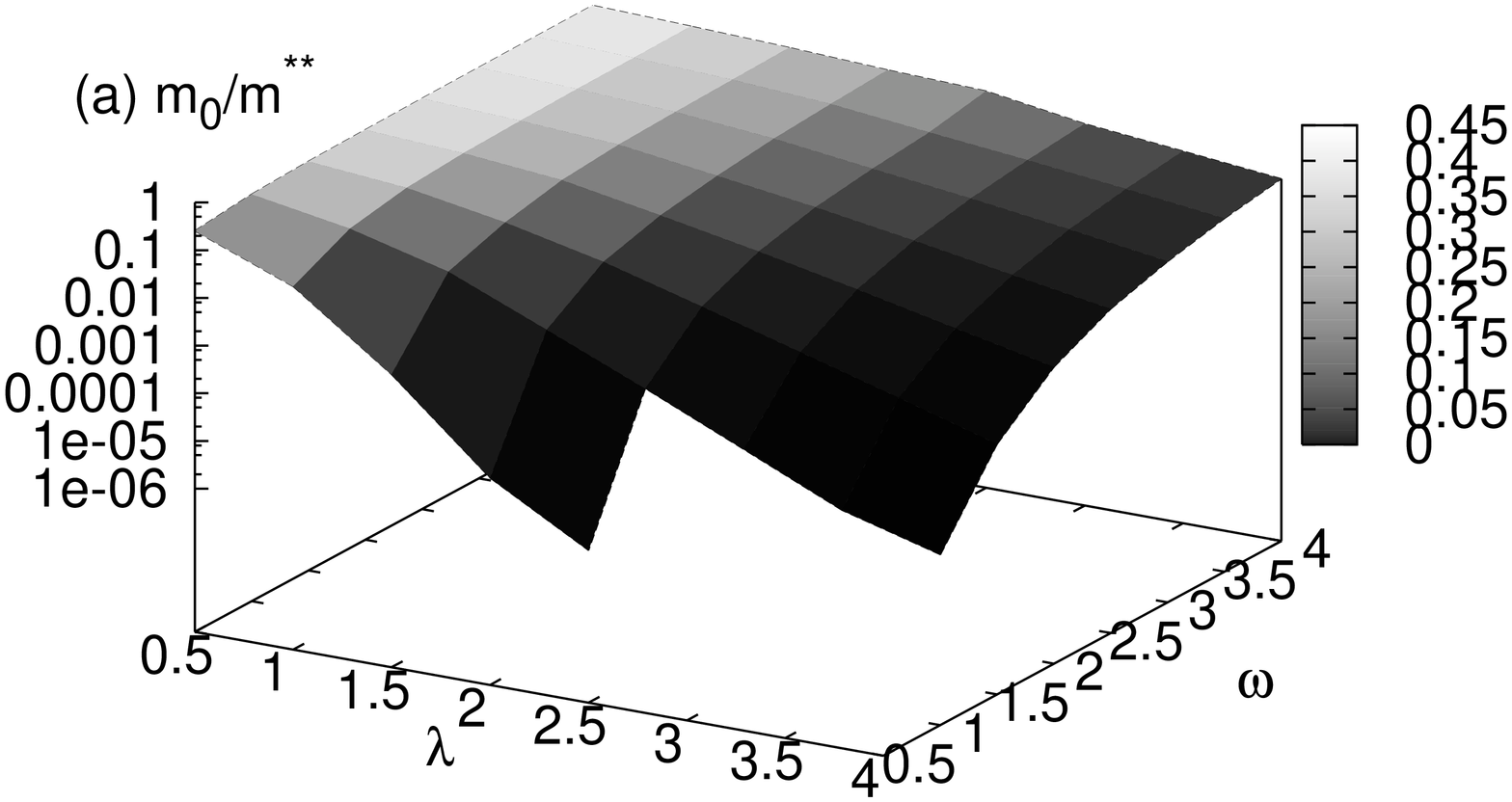}
\includegraphics[height = 65mm, angle = 270]{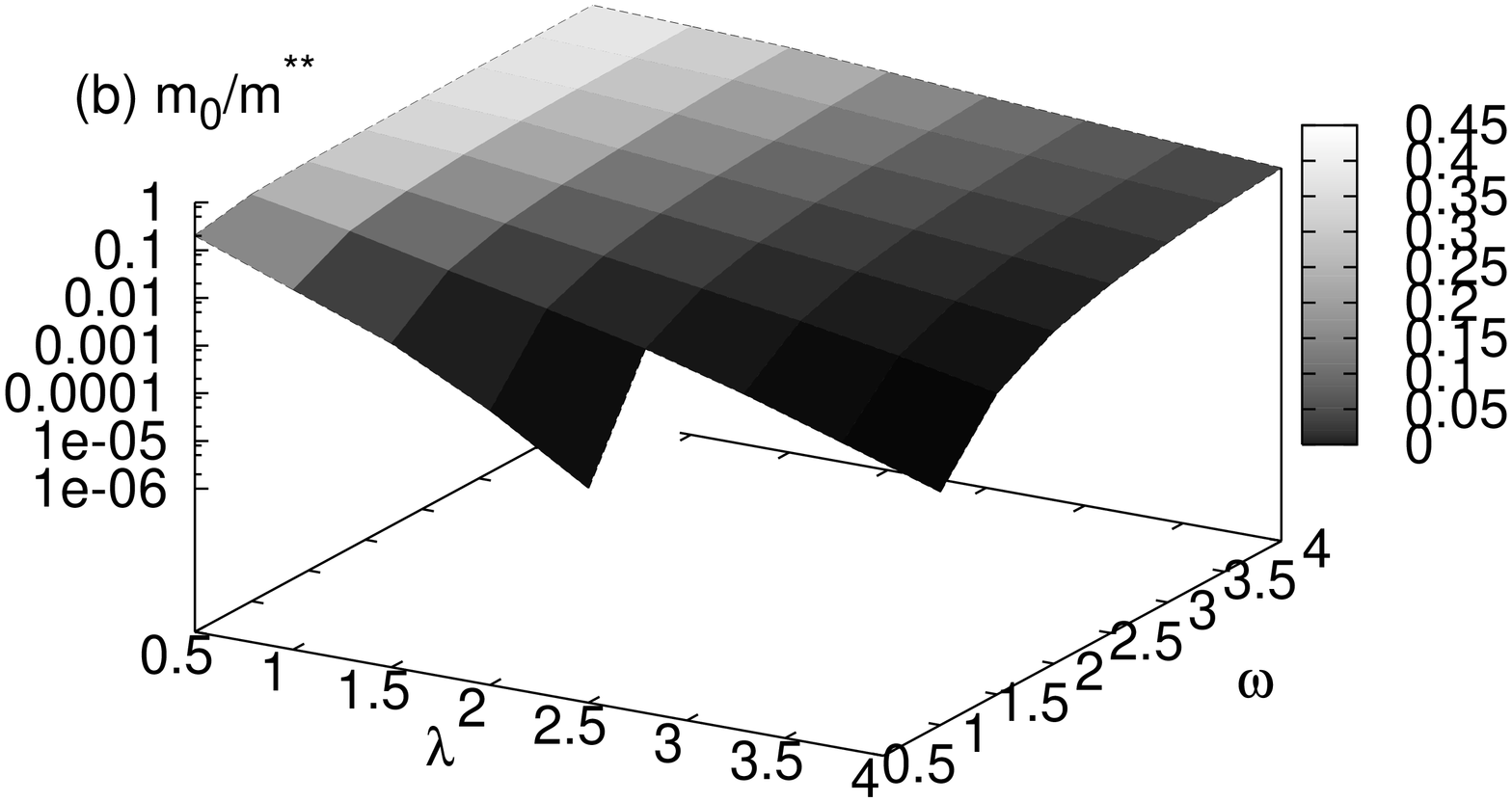}
\end{indented}
\caption{Inverse mass of the bipolaron for (a) the rectangular ladder
and (b) the staggered ladder. There is more than an order of magnitude difference between the mass of the bipolaron on the staggered ladder, and that on the rectangular ladder over significant regions of the parameter space. Bipolaron masses on the staggered ladder have recently been shown to have a value commensurate with the polaron mass.}
\label{fig:inversemass}
\end{figure}

If one is to reach a superconducting state via the Bose-Einstein
condensation of bipolarons, there are two conditions. First, the
bipolaron pair must be light, and second, the pairing radius must be
small. We demonstrate the differences between the inverse masses on
the staggered and rectangular ladders in figures
\ref{fig:inversemass}(a) and (b), which show the inverse mass of the
bipolaron for a number of different $\lambda$ and $\omega/t$. There is
more than an order of magnitude difference between the mass of the
bipolaron on the staggered ladder, and that on the rectangular ladder
over significant regions of the parameter space. In fact, the
magnitude of the bipolaron mass turns out to be of similar size to
that of the polaron mass over a wide range of the parameter space. The
mass is inversely proportional to the transition temperature of the
BEC, so a small mass is essential to obtain a decent $T_C$.

\begin{figure}
\begin{indented}\item[]
\includegraphics[height = 65mm, angle = 270]{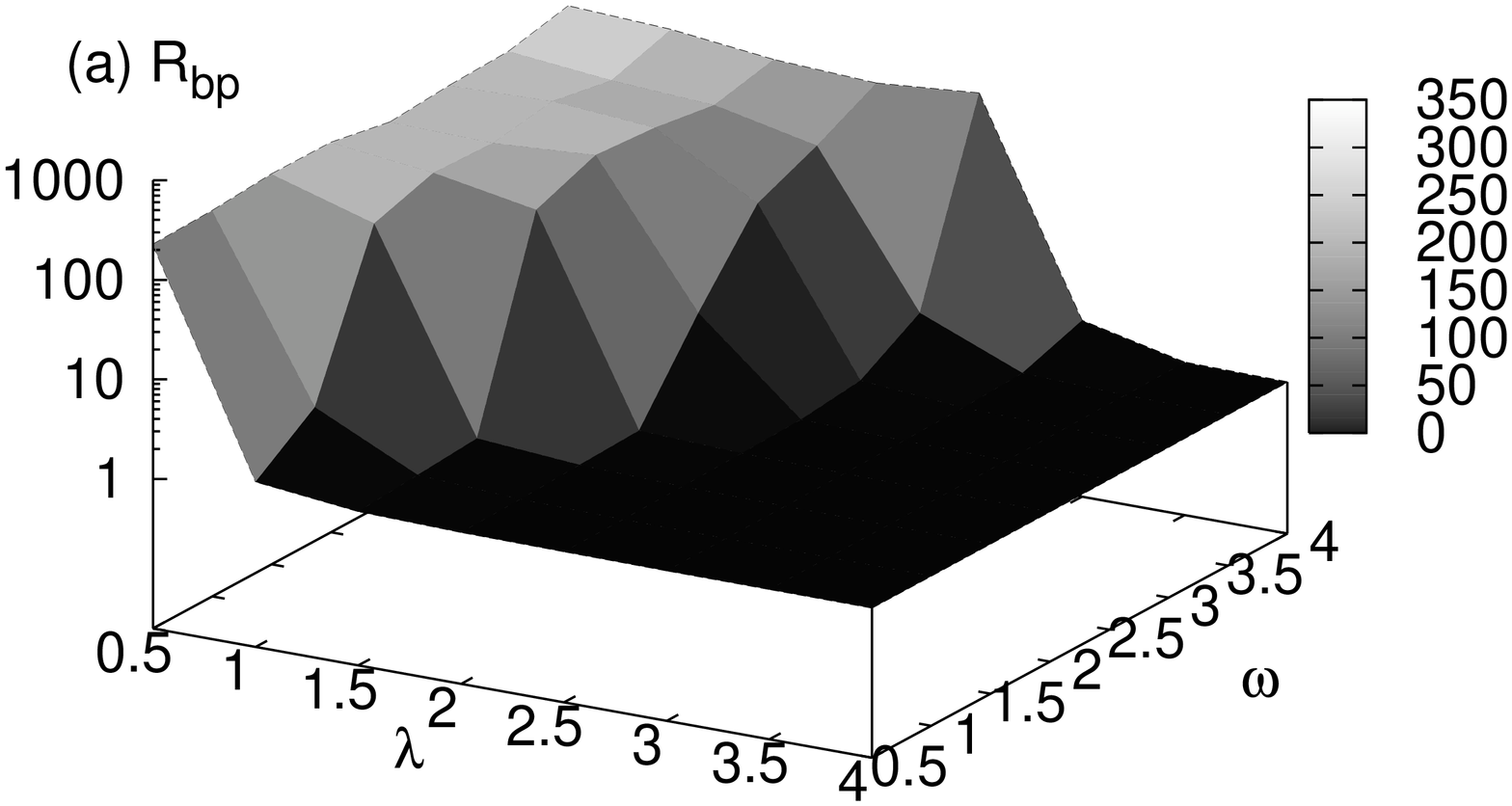}
\includegraphics[height = 65mm, angle = 270]{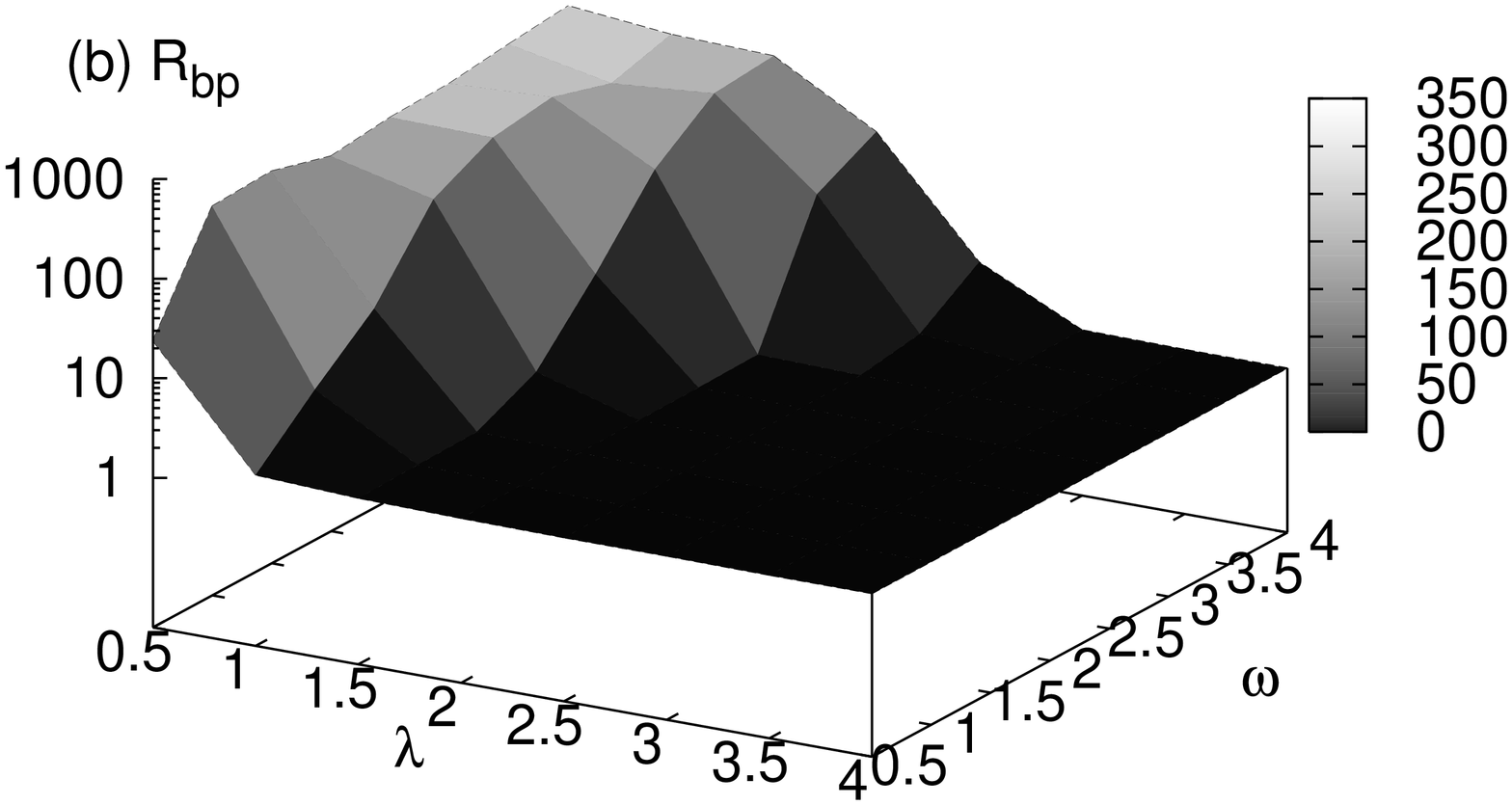}
\end{indented}
\caption{Size of the bipolaron for (a) the rectangular ladder
and (b) the staggered ladder. Not only is the bipolaron on the staggered ladder lighter than that on the rectangular ladder, its size is also smaller for equivalent $\lambda$.}
\label{fig:bipolaronsize}
\end{figure}

We have demonstrated that one of the precursors for a Bose-Einstein
condensate of pairs above the mK range may be met on the staggered
ladder arrangement, but we also require small pairs, with
non-overlapping wavefunctions. In figure \ref{fig:bipolaronsize} we
show how the the size of the bipolaron varies as $\lambda$ and
$\omega/t$ are varied. Not only is the bipolaron on the staggered
ladder lighter than that on the rectangular ladder, it is also has a
significantly smaller radius than the bipolaron on the rectangular
ladder, making it a much better prospect for Bose-Einstein
condensation.

\begin{figure}
\begin{indented}\item[]
\includegraphics[height = 65mm, angle = 270]{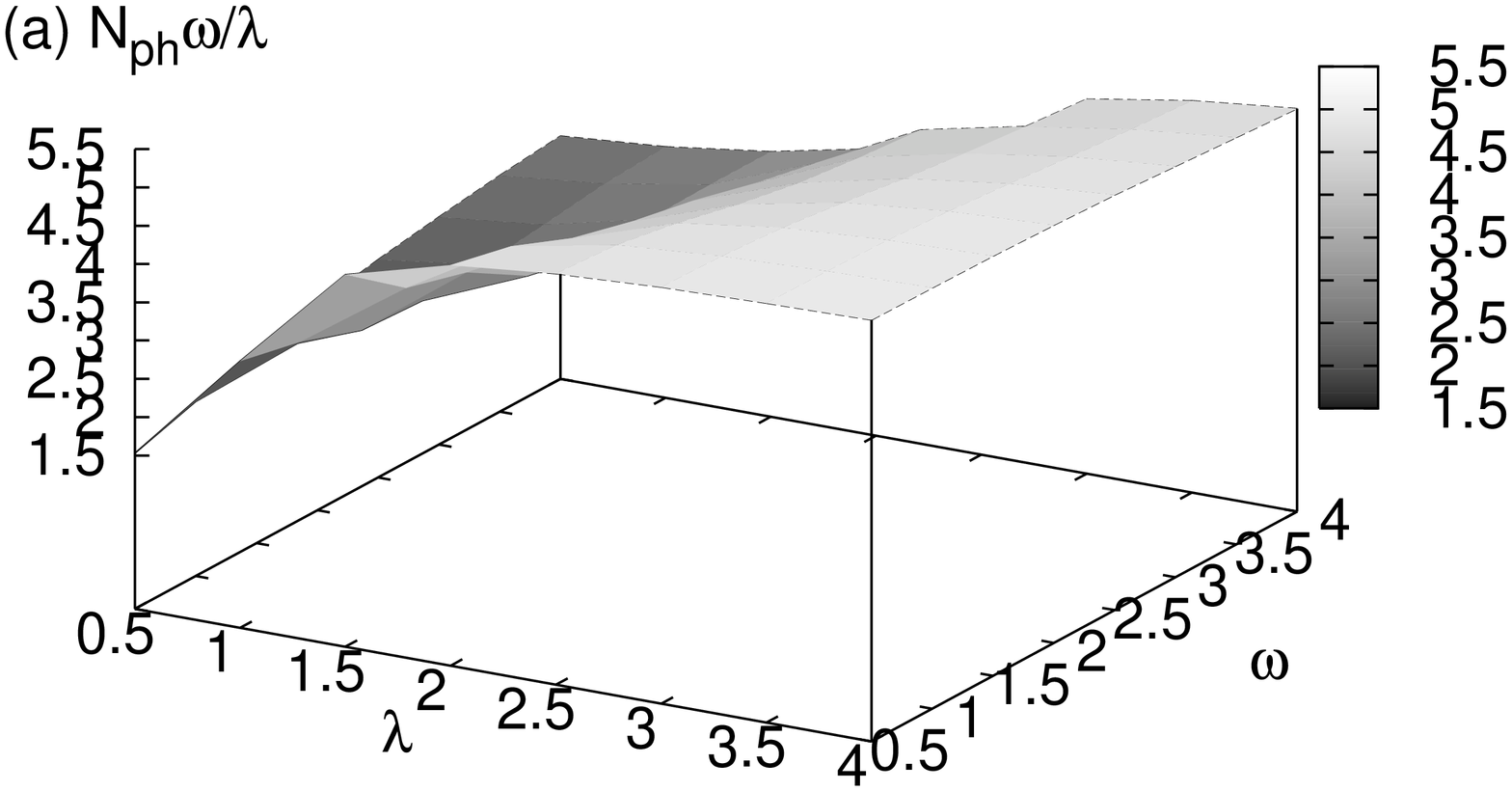}
\includegraphics[height = 65mm, angle = 270]{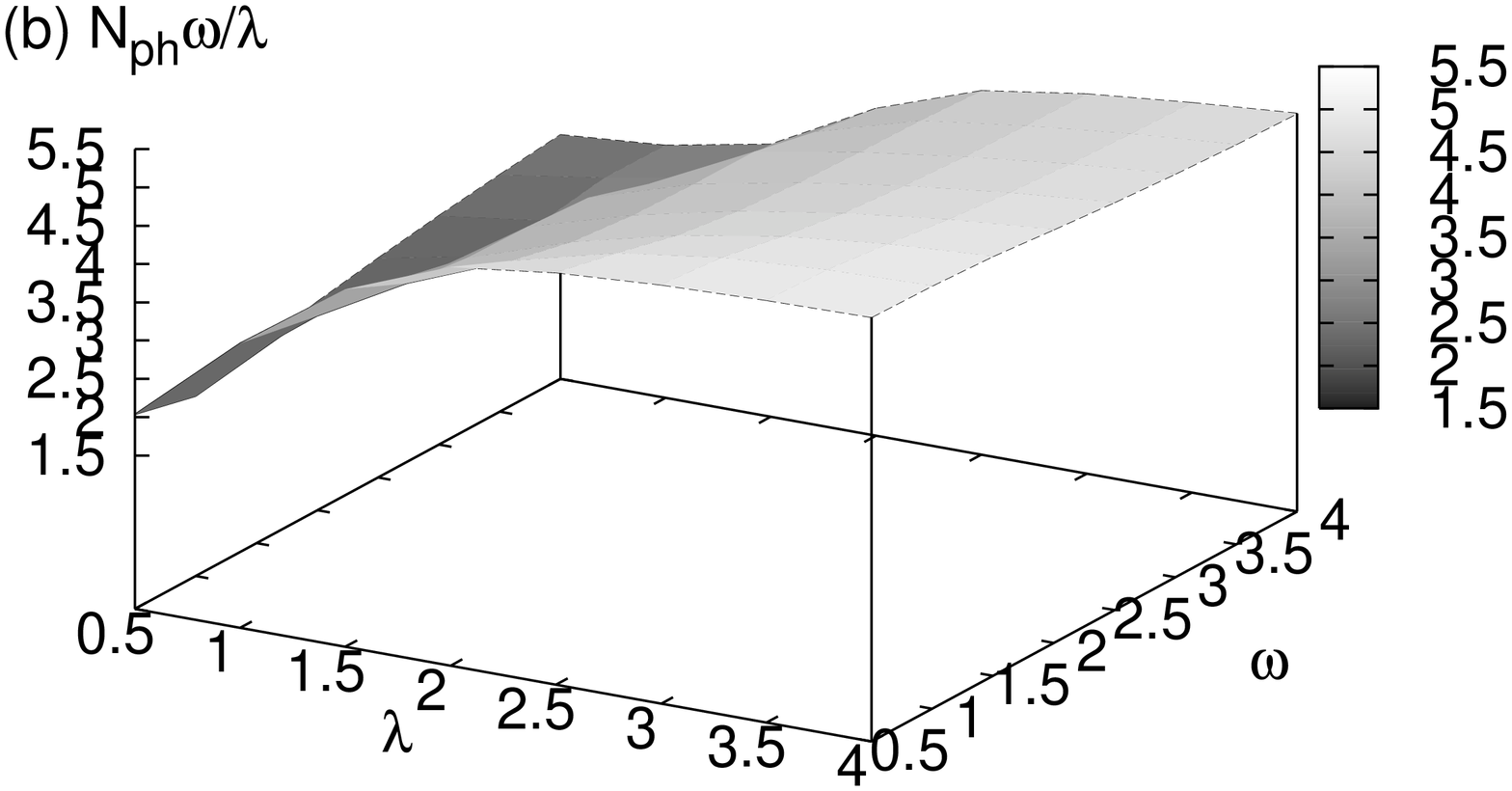}
\end{indented}
\caption{Number of phonons associated with the bipolaron for (a) the rectangular ladder
and (b) the staggered ladder, weighted by
phonon frequency and electron phonon coupling. Again, we can see that
the strong coupling limit is achieved at significantly lower $\lambda$
on the staggered ladder. It is interesting to note that the states with large omega have a smaller number of phonons. This is consistent with the anti-adiabatic approximation: When phonon frequency is large, creating phonons becomes difficult, and phonon wavefunction is the vaccum state. Then the phonon problem maps onto a $UV$ model (discussed later).}
\label{fig:nophonons}
\end{figure}

Since the result that the bipolaron mass is proportional to the
polaron mass via a numerical value relies on the anti-adiabatic
limit, where the phonon frequency is very large, so that phonons may
not be excited, we investigate the number of phonons asociated with
the bipolaron in figure \ref{fig:nophonons}. The result is weighted by
phonon frequency and electron phonon coupling, so that the strong
coupling result can clearly be seen. Again, we can see that the
strong-coupling limit is reached at significantly lower $\lambda$ on
the staggered ladder than on the rectangular ladder. Strong coupling
results from the previous section agree well, with the ratio $\omega
N_{ph}/\lambda$ approaching the numerical value given in table
\ref{tab:strong}. The anti-adiabatic limit can clearly be identified
as regions on the graph where the phonon number approaches zero
(i.e. the $\omega>\lambda$ quadrant of the parameter space),
consistent with results for the mass.

\begin{figure}
\begin{indented}\item[]
\includegraphics[height = 65mm, angle = 270]{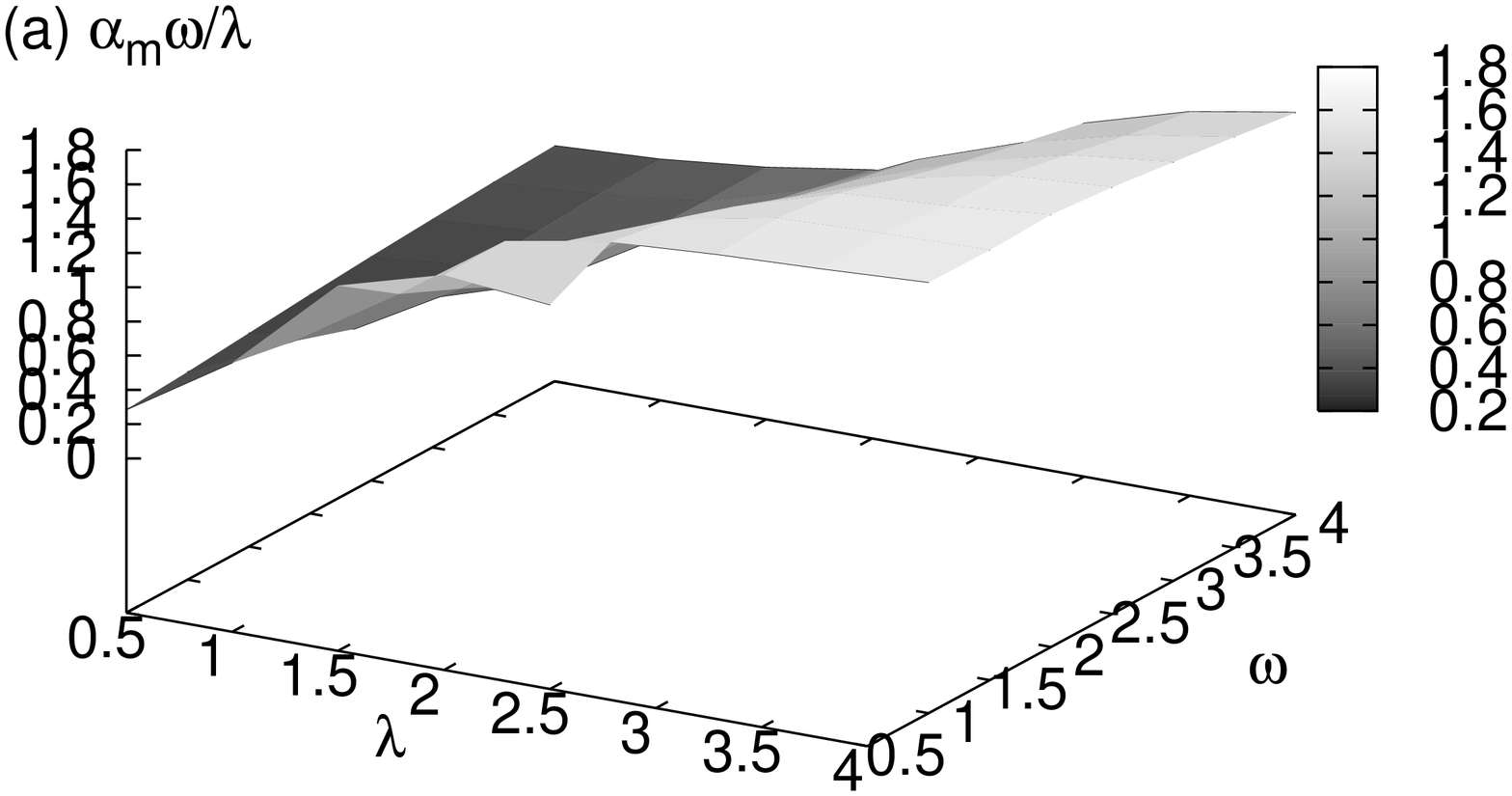}
\includegraphics[height = 65mm, angle = 270]{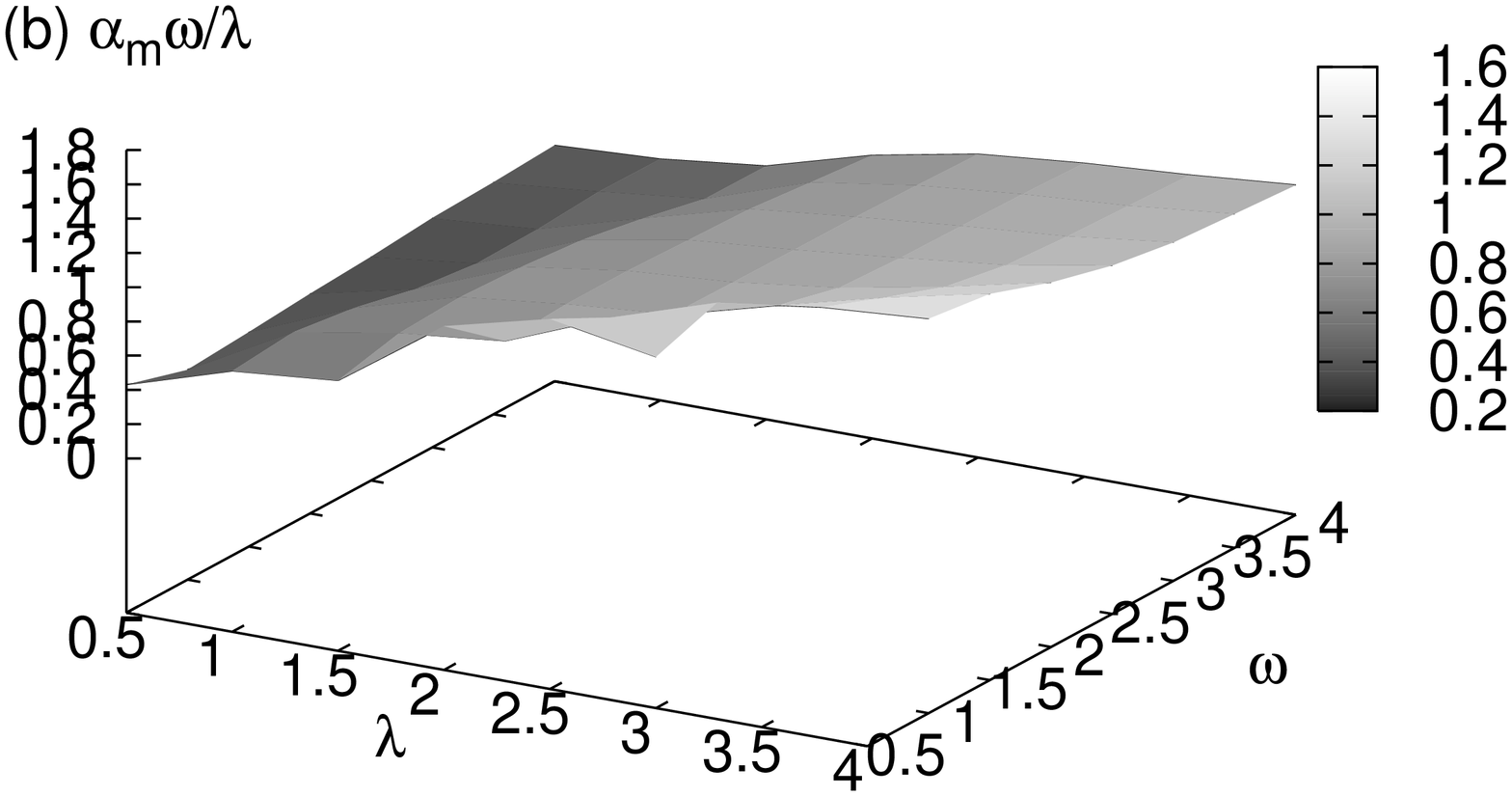}
\end{indented}
\caption{Mass isotope exponent of the bipolaron for (a) the rectangular ladder
and (b) the staggered ladder. Again, we can see that the
strong coupling limit is achieved at significantly lower $\lambda$ on
the staggered ladder. The isotope exponent is also smaller on the
staggered ladder, demonstrating a much smaller range of mass from weak
to strong coupling.}
\label{fig:iex}
\end{figure}

We show the mass isotope exponent of the bipolaron in figure
\ref{fig:iex}. Again, we can see that the strong coupling limit is
achieved at significantly lower $\lambda$ on the staggered ladder,
compared with the rectangular ladder. The isotope exponent is also
smaller on the staggered ladder, demonstrating a much smaller range of
mass from weak to strong coupling.

\begin{figure}
\begin{indented}\item[]
\includegraphics[height = 65mm, angle = 270]{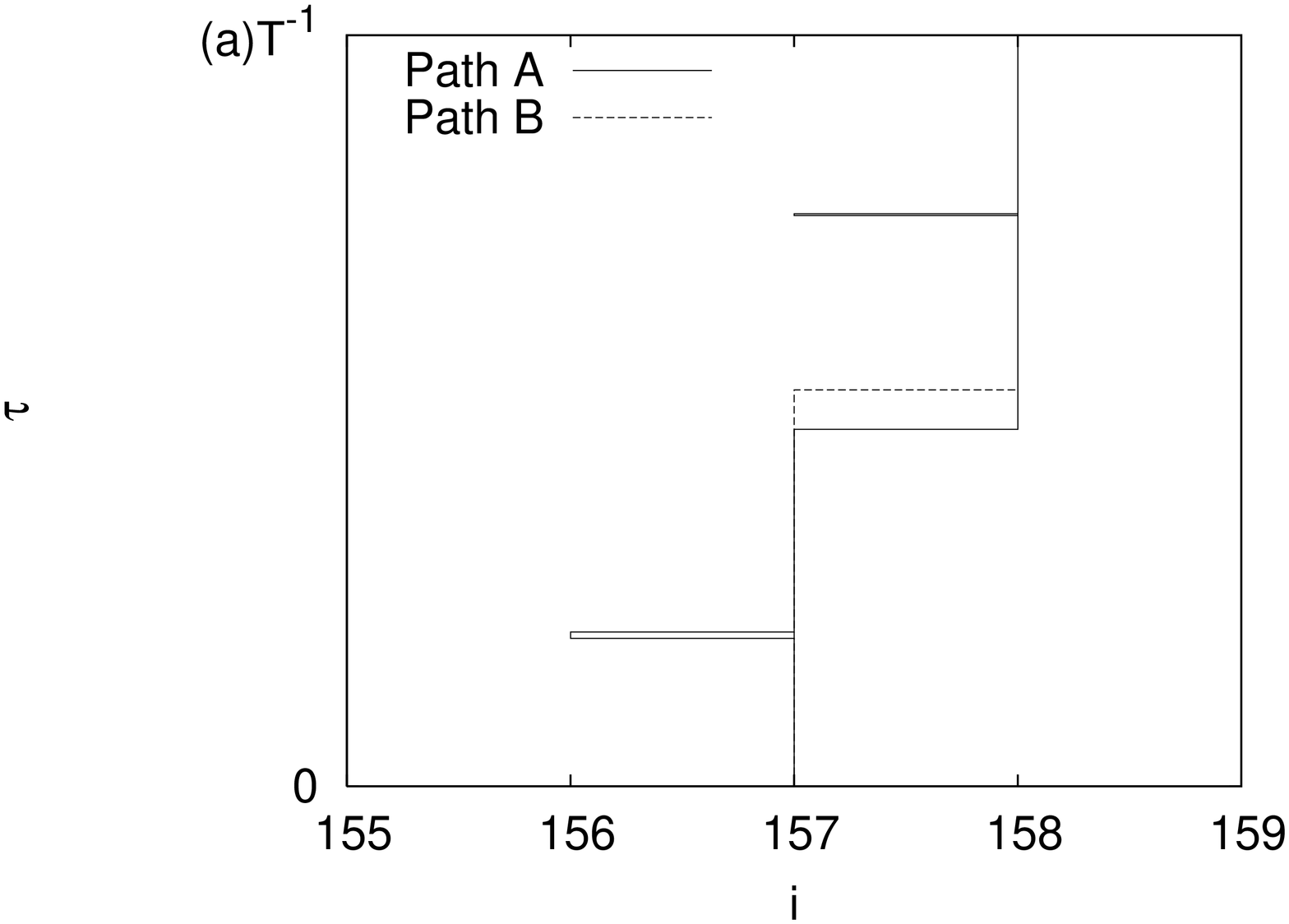}
\includegraphics[height = 65mm, angle = 270]{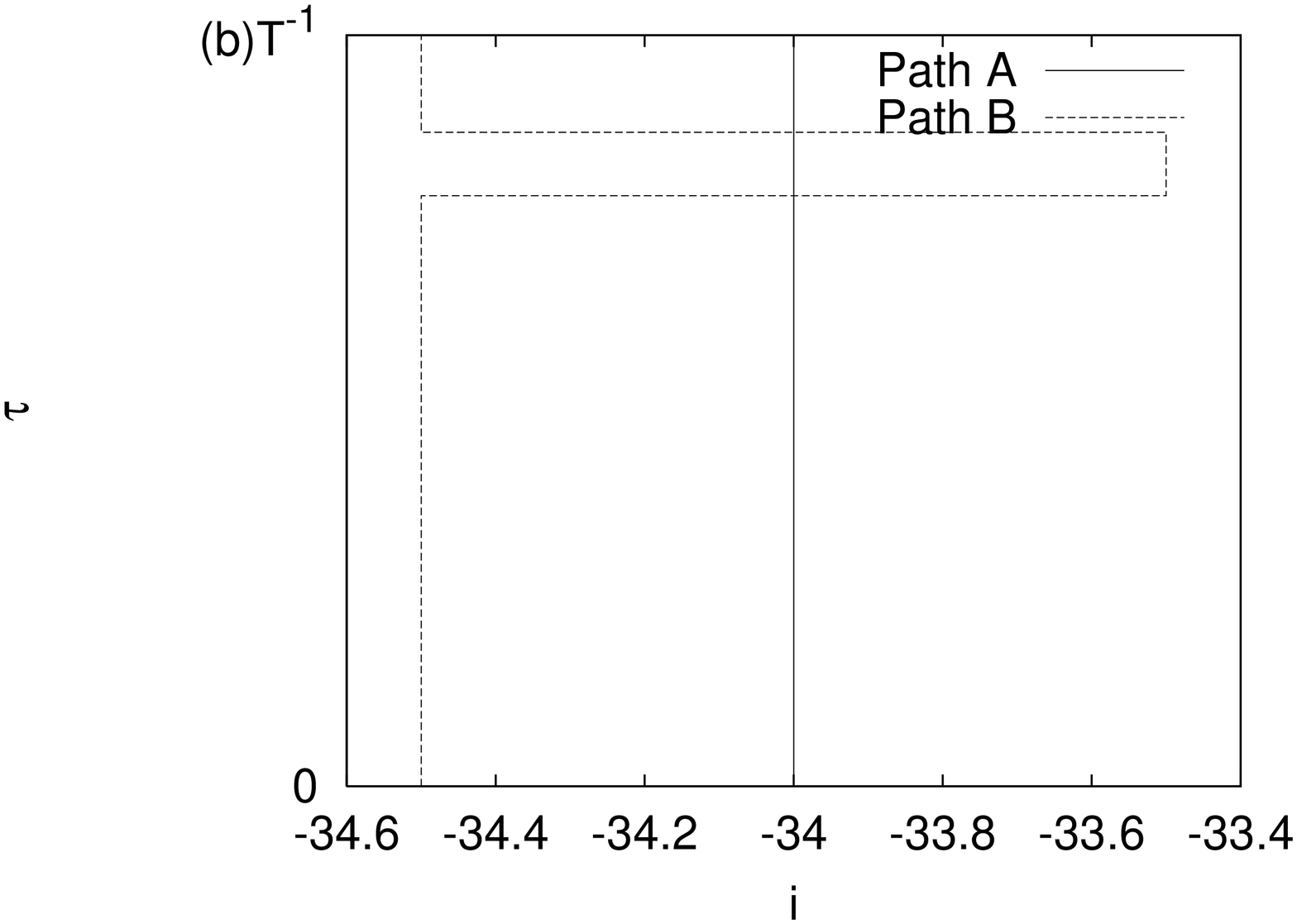}
\end{indented}
\caption{Example path configurations on (a) the rectangular ladder and (b) the staggered ladder. Hopping events on different paths on the rectangular ladder are very correlated in time. On the staggered ladder, there
are two degenerate configurations, and paths are just as likely to sit
on either of the two neighbouring sites, significantly reducing the
correlation between kinks. ($\omega/t = \lambda = 4$ in both cases.)}
\label{fig:path}
\end{figure}

Finally, in figure \ref{fig:path}, we show example path configurations
on (a) the rectangular ladder and (b) the staggered ladder. Hopping
events on different paths on the rectangular ladder are very closely
correlated on the imaginary time axis. On the staggered ladder, there
are two degenerate configurations, and paths are just as likely to sit
on either of the two neighbouring sites, significantly reducing the
correlation between kinks, and increasing the probability that kink
pairs can be inserted. It is this that leads to significantly lighter
bipolarons on the staggered ladder.

\section{Beyond ladders: Other superlight systems}

On the ladder systems, the electrons were held on neighbouring legs of
the ladders. This is partially consistent with a very strong local
Coulomb repulsion or Hubbard $U$. It is also possible to examine the
effects of a very strong or even infinite Hubbard $U$ on the masses of
bipolarons bound via long range attraction on other low dimensional
systems. As we have seen by examining the ladder systems, there are 3
requirements for superlight bipolarons (1) Electrons are not allowed
to bind on a single site (2) There are at least two degenerate
configurations of electrons sitting on neighbouring sites and (3)
There are single hopping events which transform one configuration into
another degenerate configurations. Then, hopping of the bipolaron is first
order in the hopping of the polaron. Condition (2) is satisifed by long range electron-phonon intereraction, and there are several tight binding
lattices which also satisfy conditions (2) and (3), and a strong
site-local Coulomb repulsion satisfies condition (1). We discuss
bipolarons on these lattices in the anti-adiabatic approximation in
this section.

\subsection{Triangular molecule}

The staggered ladder system discussed earlier in this paper can be
considered to be made up of triangular plaquettes, so the first logical step to looking beyond the ladder systems is to analyse the physics of a single plaquette. If hopping
is allowed between all the sites on the plaquette, then one may
consider exchange effects, which lead to singlet and triplet bound
states. To see exchange effects, we require lattices with odd-membered
rings.  As the simplest example, consider three sites $0,1,2$, all of
which are neighbours, with hopping $\tilde{t}>0$.  The polaron
Hamiltonian is
\begin{equation}
\label{tripol}
H = \left(\begin{array}{ccc}0 & -\tilde{t} & -\tilde{t} \\-\tilde{t} & 0 & -\tilde{t} \\-\tilde{t} & -\tilde{t} & 0\end{array}\right).
\end{equation}
The dimer lattice is constructed by placing a node at the centre of each bond.  The singlet bipolaron Hamiltonian is therefore 
\begin{equation}
\label{trisin}
H_S = \left(\begin{array}{ccc}V_\mathrm{min} & -\tilde{t} & -\tilde{t} \\-\tilde{t} & V_\mathrm{min} & -\tilde{t} \\-\tilde{t} & -\tilde{t} & V_\mathrm{min}\end{array}\right)
\end{equation}
with eigenvalues $\{V_\mathrm{min}-2\tilde{t},V_\mathrm{min}+\tilde{t},V_\mathrm{min}+\tilde{t}\}$, and the triplet bipolaron Hamiltonian is
\begin{equation}
\label{tritri}
H_T = \left(\begin{array}{ccc}V_\mathrm{min} & \tilde{t} & \tilde{t} \\t & V_\mathrm{min} & \tilde{t} \\t & \tilde{t} & V_\mathrm{min}\end{array}\right)
\end{equation}
with eigenvalues
$\{V_\mathrm{min}-\tilde{t},V_\mathrm{min}-\tilde{t},V_\mathrm{min}+2\tilde{t}\}$.
(An alternative choice of basis would be the six-dimensional
unsymmetrised $S_z=0$ basis $\{|i\uparrow j\downarrow\rangle\}$.  This
would transform the triangle into a six-member ring.  Such a basis
becomes unwieldy for typical lattices). Clearly, the properties of the
plaquette are defined by single polaron hopping events, since there
are no eigenvalues with $O(\tilde{t}^{2})$ terms. It is this property
which makes lattices which can be constructed from triangular
plaquettes a good starting place when looking for small superlight
bipolarons.

\subsection{Triangular lattice}

Let us consider a triangular lattice with nearest-neighbour hopping $t$. In the anti-adiabatic approximation, this is replaced by $\tilde{t}$. The polaron band structure (ignoring polaron shift) is
$E(\mathbf{k}) = -2\tilde{t} C(\mathbf{k})$, where we have defined
\begin{eqnarray}
\label{C}
C(\mathbf{k})\equiv\cos k_x a & + & \cos\left(k_x a/2 -\sqrt{3} k_y a/2\right)\nonumber\\
& & + \cos\left(k_x a/2 +\sqrt{3} k_y a/2\right),
\end{eqnarray}
$a$ being the lattice parameter.  Expanding near the $\Gamma$ point gives 
\begin{equation}
\label{Epol0}
E(\mathbf{k}) = -6\tilde{t} + \frac32 k^2 a^2 \tilde{t} + O(k^4)
\end{equation}
The polaron effective mass is therefore
\begin{equation}
\label{mpol}
m^* = \frac{\hbar^2}{3\tilde{t}a^2}.
\end{equation}
\begin{figure}
\begin{indented}\item[]
\includegraphics[width=12cm]{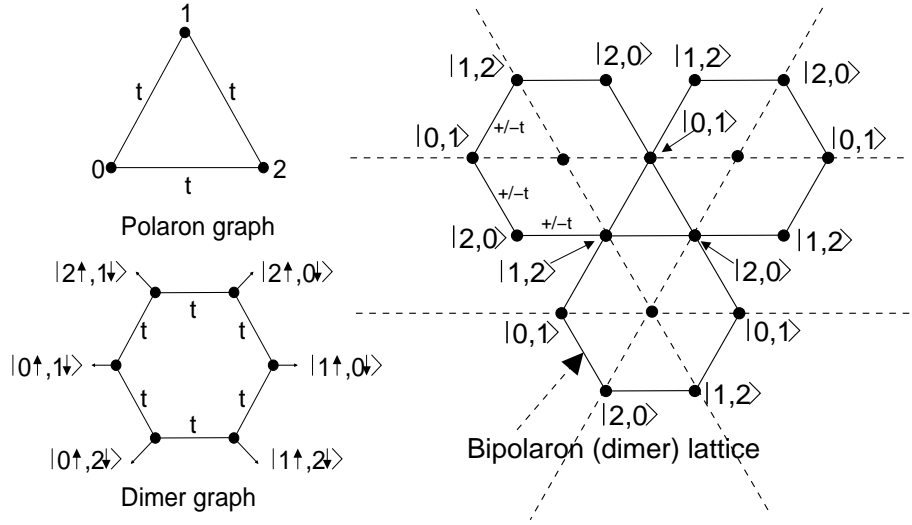}
\caption{The triangular lattice (thin lines) with a dimer state (circle) at the mid-point of each bond.  The polaron and bipolaron states are indicated on one triangle, with spin indices suppressed.  The dimer lattice (thick lines) is a kagome lattice.}
\end{indented}
\label{kagomefig}
\end{figure}
By placing a node on each bond in the lattice, we see that the resulting dimer lattice is a kagome lattice.  The dispersion is easily found as follows (see also ref \cite{xpch}) by diagonalising  the secular matrix
\begin{equation}
\label{Hkagome}
H(\mathbf{k}) =V_\mathrm{min}\mp \tilde{t} \left|\begin{array}{ccc}0 & 1+\gamma^* & 1+\beta \\1+\gamma & 0 & 1+\alpha^* \\1+\beta^* & 1+\alpha & 0\end{array}\right|,
\end{equation}
with
\begin{eqnarray}
\alpha & = & \exp\left(-ik_x a/2 +i \sqrt{3} k_y a/2\right)\\
\beta & = & \exp\left(-ik_x a/2 -i \sqrt{3} k_y a/2\right)\\
\gamma & = & \exp\left(ik_x a\right).
\end{eqnarray}
The sign in (\ref{Hkagome}) is $-\tilde{t}$ for the singlet and $+\tilde{t}$ for the triplet.
There are three bipolaron bands with no gaps:
\begin{eqnarray}
E_1(\mathbf{k}) & = & V_\mathrm{min}\pm \tilde{t} \left(-1 - \sqrt{3+2C(\mathbf{k})}\right) \label{band1} \\
E_2(\mathbf{k}) & = & V_\mathrm{min}\pm \tilde{t} \left(-1 + \sqrt{3+2C(\mathbf{k})}\right) \label{band2} \\
E_3(\mathbf{k}) & = &V_\mathrm{min} \pm 2\tilde{t}.\label{band3}
\end{eqnarray}
with sign $+\tilde{t}$ for the singlet and $-\tilde{t}$ for the triplet.

The lowest singlet band is 
\begin{equation}
\label{E1s}
E_1(\mathbf{k}) = V_\mathrm{min} - 4\tilde{t} + \frac14 k^2 a^2 \tilde{t} + O(k^4),
\end{equation}
with effective mass 
\begin{equation}
\label{mbpol}
m_s^{**} =  \frac{2\hbar^2}{\tilde{t}a^2} = 6m^*.
\end{equation}
We notice that this increases \emph{linearly} with the polaron mass, indicating that crab-like bipolarons can be relatively light.

The lowest triplet band is the \emph{flat band}
$E_3(\mathbf{k})=V-2\tilde{t}$.  This implies that a triplet crab-like
bipolaron has \emph{infinite} mass on a triangular lattice.  Once
crawler motion is permitted, the effective mass is expected to be
finite but proportional to $(m^*)^2$.

\subsection{Lattices with long range hopping}

Lattices with triangular components are not the only ones where the
ability to move between degenerate paired states can be utilised to
lead to small masses. For example, if we introduce
next-nearest-neighbour hopping to a linear chain, then, in the
anti-adiabatic approximation, we obtain a bipolaron mass,
$2(2a)^2\tilde{t}'$ and a polaron mass $2a^2\tilde{t}$ where
$\tilde{t} = t\exp(-W\lambda(1-\Phi(0,{\bf a})/\Phi(0,0))/\omega)$ and
$\tilde{t}' = t'\exp(-W\lambda(1-\Phi(0,2{\bf
a})/\Phi(0,0))/\omega)$. While this does not lead to an exact
cancellation of the exponents, the bipolaron to polaron mass ratio is
linear in $t'/t$, and the bipolaron is therefore expected to be light.

Another suggestion for bipolarons with light mass comes from the
cuprate lattice. It has often been suggested that the tight-binding
structure of the cuprates is a square lattice with nearest-neighbour
hopping $t$ and next-nearest-neighbour hopping $t'$. In the presence
of very stong Coulomb repulsion $U$, one may determine that the ground
state of the Lang-Firsov transformed Hamiltonian consists of nearest
neighbour pairs with degeneracy four. If one applies the
anti-adiabatic approximation to such a lattice, one determines that
the hopping of such a bipolaron is first order in $t'$. We obtain a
bipolaron mass, $2(\sqrt{2}a)^2\tilde{t}'$ and a polaron mass
$2a^2\tilde{t}$ where $\tilde{t} = t\exp(-W\lambda(1-\Phi(0,{\bf
a})/\Phi(0,0))/\omega)$ and $\tilde{t}' =
t'\exp(-W\lambda(1-\Phi(0,{\bf
a}-R^{90}{\bf{a}})/\Phi(0,0))/\omega)$. Here $R^{90}$ is the rotation
operator. Again, while we don't get an exact cancellation of the
exponents, we obtain a resulting mass proportional to $t'/t$, and
light bipolarons are likely \footnote{It is interesting to note that
these lattices share a common feature with triangular lattices, namely
that it is possible for a single particle to hop 3 times to return to
the point of origin, indicating a more universal prospect for this
phenomenon}. Such pairing is an appealing prospect, especially since
the Hague has demonstrated the potential for d-wave superconductivity
mediated by Holstein-like electron-phonon interactions in the
intermediate coupling limit \cite{hague2006a}.

\begin{figure}
\begin{indented}\item[]
\includegraphics[width = 11cm]{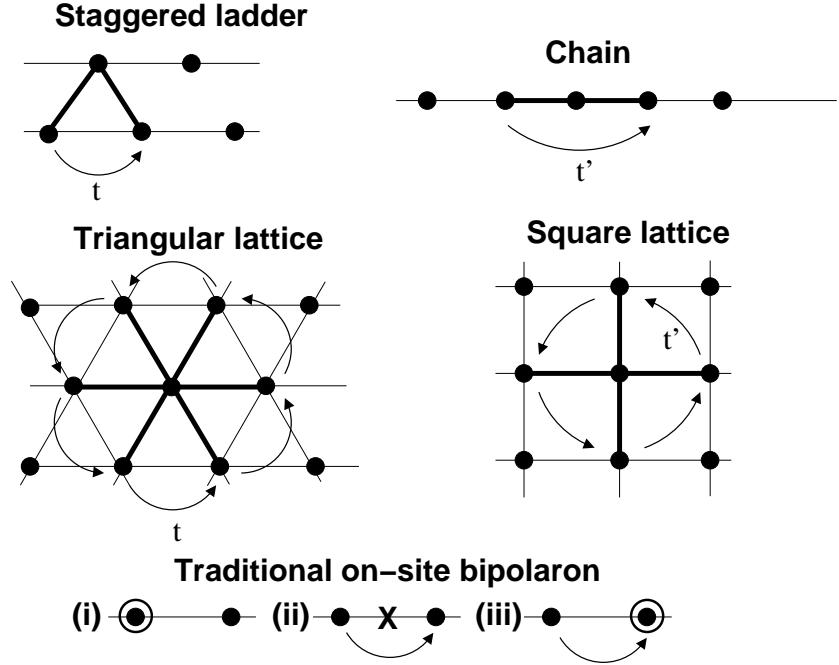}
\end{indented}
\caption{Pictorial demonstration of lattices which are expected to
lead to superlight and light bipolarons when long range
electron-phonon interaction is present in combination with a strong
site-local Coulomb repulsion (Hubbard U) which stops on-site
binding. All degenerate nearest-neighbour bound states are shown as
the thick lines. The lightest bipolarons are expected on lattices with
a trangular component, where hopping between degenerate states can be
realised with the same single hop as the polaron (shown as
arrows). Thus the bipolaron moves in the first order of the polaron
mass. We also show potential single hops for light bipolarons on the
chain, and the square lattice, which can be realised if
next-nearest-neighbour hopping is present. The superlight bipolarons
are contrasted with traditional (Holstein) on-site bipolarons, which
move with two hopping events, the first one breaking the bound state
(step ii), and the second reforming it (step iii), thus such on-site
(crawler) bipolarons cannot be simultaneously small (well bound) and
mobile.}
\label{fig:pictorial}
\end{figure}

\begin{figure}
\begin{indented}\item[]
\includegraphics[height = 9cm, angle=270]{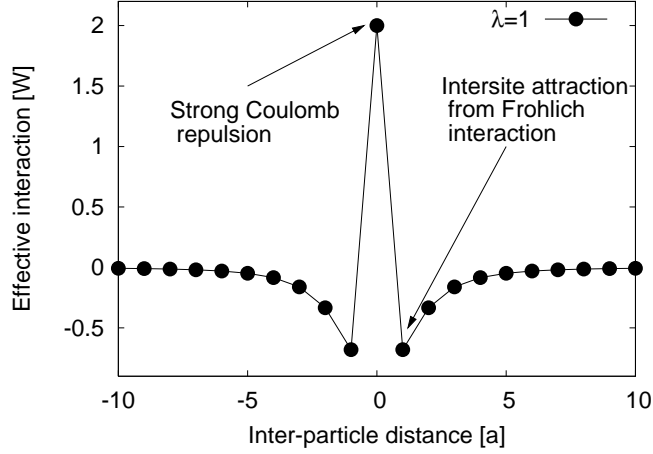}
\end{indented}
\caption{Pictorial demonstration of the interpolaron effective interaction which can lead to superlight bipolarons (cross-section of anti-adiabatic limit of triangular lattice with $\lambda=1$). A strong on-site Coulomb repulsion (Hubbard $U$) stops on-site binding, leading to binding between polarons on neighbouring sites when there is a long range phonon mediated attraction. In this case, the potential well should lead to binding of small bipolarons on the order of one lattice site. The long range Fr\"ohlich electron-phonon attraction may be found on quasi-2D lattices, where ions oscillate above planes (as described in reference \cite{alekor}.}
\label{fig:effinteraction}
\end{figure}

\section{Conclusions}

 The bipolaronic extension \cite{alebook} of the BCS theory towards the
strong interaction between electrons and ion vibrations proved that
the Cooper pairing in momentum space \cite{bcs} and the
Ogg-Schafroth real-space pairing \cite{ogg,shaf}  are two extreme
limits of the same problem.
 For a very strong electron-phonon coupling, polarons
become self-trapped on a single lattice site and bipolarons are
on-site singlets. In the Holstein model of the zero-range EPI
 their mass  appears only in the second order of polaron hopping \cite{aleran},
 so that on-site bipolarons are very heavy. This estimate led some authors to
 the conclusion that the formation
of itinerant small polarons and bipolarons in real materials is
unlikely \cite{mel}, and high-temperature bipolaronic
superconductivity is impossible \cite{and2}.

In fact, we have demonstrated here that small but light bipolarons
could exist for realistic values of the finite-range EPI with
high-frequency optical phonons in staggered ladder systems. Small
light bipolarons are an essential precursor to high-temperature
superconductivity, since the Bose-Einstein condensate has transition
temperature that is inversely proportional to mass, and wavefunctions
may not overlap.  Such bipolarons are easily formed on lattices with
triangular plaquettes in the presence of extremely large on-site
Coulomb repulsion, and persist to large EPI. This conclusion is backed
up by analytics in the anti-adiabatic approximation in the presence of
large intersite Coulomb attraction. Another important conclusion is
that the triplet-singlet exchange energy is of the first order in the
hopping integral, and triplet bipolarons are heavier than singlets in
certain lattice structures at variance with simple intuitive
expectations. We summarise the types of lattices where light ``crab''
bipolarons may be formed in figure \ref{fig:pictorial}, contrasting
with the traditional Holstein bipolarons (bottom) and describe the
required effective interaction in figure \ref{fig:effinteraction}
demonstrating the underlying physics of such bipolarons.

Our CTQMC simulations lead us to believe that the following recipe
is worth investigating to look for very high-temperature
superconductivity: (a) The parent compound should be an ionic
insulator with light ions to form high-frequency optical phonons,
(b) The structure should be quasi two-dimensional to ensure poor
screening of high-frequency c-axis polarized phonons, (c) A
triangular lattice is required in combination with  strong, on-site
 Coulomb repulsion to form the small superlight Crab
bipolaron  (d) Moderate carrier densities are required to keep the
system of small bipolarons close to the dilute regime. Many
of these conditions are already met in the cuprates.

\section{Acknowledgements}

We would like to acknowledge EPSRC (grant numbers EP/C518365/1 and
EP/D07777X/1). While writing this article, conversations with Serge
Aubry, Peter Edwards, Janez Bon\v{c}a, Jozef Devresse, Holger Fehske,
Yurii Firsov, Martin Hohenadler, Viktor Kabanov, Wolfgang von der
Linden, Peter Littlewood, Dragan Mihailovic, Arndt Simon, Marshall
Stoneham, and other participants of the European Science Foundation
workshop ``Mott's Physics in Nanowires and Quantum Dots'' (Cambridge,
UK, 31 July-2 August, 2006) were especially helpful.  We acknowledge
support from Loughborough University, Gonville and Caius College,
Cambridge University, and the ESF for this workshop.


\begin{thebibliography}{199}

\bibitem{bcs} J. Bardeen, L. N. Cooper, and J. R. Schrieffer, Phys. Rev. {\bf 108}, 1175 (1957).

\bibitem{alemott94} A. S.  Alexandrov and N. F. Mott, Rep. Prog. Phys. ${\bf 57}$,
1197 (1994).

\bibitem{aleedwards}
A. S. Alexandrov and P. P. Edwards, Physica C (Amsterdam) {\bf 331},
97 (2000).

\bibitem{alebook} A. S. Alexandrov, \emph{Theory of
Superconductivity: From Weak to Strong Coupling} (IoP Publishing,
Bristol, 2003).

\bibitem{edwards}
P. P. Edwards, C. N. R. Rao, N. Kumar, and A. S.  Alexandrov,
ChemPhysChem {\bf 7}, 2015 (2006).

\bibitem{ale96} A. S. Alexandrov, Phys. Rev. B {\bf 53}, 2863
(1996).

\bibitem{tJ} P. W. Anderson, P. A. Lee, M. Randeria, T. M. Rice, N. Trivedi, and F. C.
Zhang, J. Phys.: Condens. Matter {\bf 16}, R755 (2004).

\bibitem{plasma} J.H. Kim, B.J. Feenstra, H.S. Somal, D. van der
Marel, W.Y. Lee, A.M. Gerrits, and A. Wittlin,  Phys. Rev. B{\bf
49}, 13065 (1994).

\bibitem{alekor}  A. S. Alexandrov and P. E. Kornilovitch, J.
Phys.: Condens. Matter {\bf 14}, 5337 (2002).


\bibitem{guo}
G. Zhao and D. E. Morris, Phys. Rev. B {\bf 51}, 16487 (1995); G.-M.
Zhao, M. B. Hunt, H.  Keller, and K. A. M\"uller, Nature {\bf
 385}, 236 (1997); R. Khasanov, D. G. Eshchenko, H. Luetkens, E. Morenzoni, T. Prokscha,
 A. Suter, N. Garifianov, M. Mali, J. Roos, K. Conder, and H. Keller
Phys. Rev. Lett. {\bf 92}, 057602 (2004).

\bibitem{aleiso}
A. S. Alexandrov,   Phys. Rev. B {\bf 46}, 14932 (1992).

\bibitem{lan0}
A. Lanzara, P. V. Bogdanov, X. J.  Zhou, S. A.  Kellar, D. L.  Feng,
E. D. Lu, T. Yoshida, H.  Eisaki, A. Fujimori, K. Kishio, J. I.
Shimoyana, T. Noda, S. Uchida, Z. Hussain, Z. X. Shen,  Nature
 \textbf{2001},  412, 510; G. H. Gweon,  T. Sasagawa, S. Y. Zhou,  J. Craf, H. Takagi, D. H.
Lee, A. Lanzara, Nature \textbf{2004}, 430,
 187.

\bibitem{shen} W. Meevasana, N. J. C. Ingle, D. H. Lu, J. R. Shi,  F.
Baumberger, K. M. Shen,  W. S. Lee,  T. Cuk,  H. Eisaki,  T. P.
Devereaux, N. Nagaosa,  J. Zaanen, and Z.-X. Shen, Phys. Rev. Lett.
{\bf 96}, 157003 (2006).

\bibitem{tun} J. Lee, K. Fujita, K. McElroy, J. A. Slezak,
M. Wang, Y. Aiura, H. Bando, M. Ishikado, T. Masui, J. X. Zhu, A. V.
Balatsky, H. Eisaki, S. Uchida, and J. C. Davis, Nature  {\bf 442},
546 (2006).


\bibitem{opt}  D. Mihailovic, C.M. Foster, K. Voss, and A.J. Heeger, Phys. Rev. B{\bf 42}, 7989 (1990);
  P. Calvani, M. Capizzi, S. Lupi, P. Maselli, A. Paolone, P. Roy, S.W. Cheong, W. Sadowski,
and E. Walker, Solid State Commun. {\bf 91}, 113 (1994); R. Zamboni,
G. Ruani, A.J. Pal, and C. Taliani, Solid St. Commun. {\bf 70}, 813
(1989).

\bibitem{neutron} T.R. Sendyka, W. Dmowski, and T. Egami, Phys. Rev. B{\bf 51}, 6747 (1995);
T. Egami, J. Low Temp. Phys. {\bf 105}, 791 (1996).

\bibitem{ver} G. Verbist, F. M. Peeters, and J. T. Devreese, Phys.
Rev. B {\bf 43}, 2712 (1991); Solid State Commun. \textbf{76}, 1005
(1990).


\bibitem{aleran}
A. S. Alexandrov  and J. Ranninger, Phys. Rev. B {\bf 23}, 1796
(1981), ibid {\bf 24}, 1164 (1981). Under certain conditions
relatively mobile intersite bipolarons can be found  even in the
Holstein-Hubbard model, generally unfavourable for coherent
tunnelling [A. La Magna and R. Pucci, Phys. Rev. B {\bf 55}, 14886
(1997); L. Proville and  S. Aubry, Eur. Phys. J. B {\bf 11}, 41
(1999); J. Bon\v{c}a, T. Katrasnic, and S.~A. Trugman, Phys. Rev.
Lett. {\bf 84}, 3153 (2000); A. Macridin G.~A.~Sawatzky, and
M.~Jarrell, Phys. Rev. B {\bf 69}, 245111 (2004)].

\bibitem{korpolaron}
P.E.Kornilovitch, \textit{Phys. Rev. Letts} \textbf{81} 5382 (1998), \textit{Phys. Rev. B} \textbf{60} 3237 (1999)

\bibitem{spencer}
P.E.Spencer, J.H.Samson, P.E.Kornilovitch and A.S.Alexandrov, \textit{Phys. Rev. B} \textbf{71} 184310 (2005)

\bibitem{hague}
J.P.Hague, P.E.Kornilovitch, A.S.Alexandrov and J.H.Samson, \textit{Phys. Rev. B} \textbf{73} 054303 (2006)

\bibitem{haguebipolaron}
J.P.Hague, P.E.Kornilovich, J.H.Samson and A.S.Alexandrov, \textit{Phys. Rev. Lett.} \textbf{98} 037002 (2007), J.P.Hague, P.E.Kornilovich, A.S.Alexandrov and J.H.Samson, \textit{Physica C} In press (2007)

\bibitem{fir}
I. G.  Lang  and Yu. A. Firsov, \textit{Zh Eksp Teor Fiz} \textbf {43} 1843 (1962) (\textit{Sov Phys JETP} \textbf{16} 1301
(1963))

\bibitem{tyablikov}
S. V. Tyablikov, Zh. Eksp. Teor. Fiz. {\bf 21}, 377 (1951); \textit{ibid} {\bf 23},  381 (1952) (in Russian).

\bibitem{xpch} Y Xiao, V Pelletier, P M Chaikin and D A Huse, Phys Rev B \textbf{67} 104505 (2003)

\bibitem{kornilovitch_mott}
P. E. Kornilovitch, to appear in this volume (2007)

\bibitem{ogg} R. A. Ogg Jr., Phys. Rev. {\bf 69}, 243 (1946).
\bibitem{shaf} M. R. Schafroth, Phys. Rev. {\bf 100}, 463 (1955); J. M.  Blatt and S. T. Butler, Phys. Rev. {\bf %
100}, 476 (1955).

\bibitem{hague2006a}
J.P.Hague, Phys. Rev. B {\bf 73} 060503(R) (2006), J. Phys.: Condens. Matter {\bf 17} 5663 (2005), \textit{ibid} {\bf 15} 2535 (2003).

\bibitem{mel} E. V. L.  de Mello and J.
Ranninger, Phys. Rev. B {\bf 58}, 9098 (1998).

\bibitem{and2} P. W. Anderson,  \emph{The Theory of Superconductivity in the
Cuprates} (Princeton University Press, Princeton NY, 1997).

\end{thebibliography}
\end{document}